\documentclass[12pt,a4]{article} 
\usepackage{epsfig}

\usepackage{graphicx}   
\usepackage{epsfig}     

\usepackage{here}

\usepackage{psfrag}   

\usepackage{amsfonts,amsbsy,amsmath,amssymb} 

\begin{document}
\begin{titlepage}
\vspace{3 cm}
\begin{center}
\begin{huge}
{\bf Discovery potential for a charged Higgs boson decaying in the 
     chargino-neutralino channel of the ATLAS detector at the LHC}\\*[.5cm]

\end{huge}
\end{center}
\vskip 1.3 cm
\begin{center}
        \normalsize {C. Hansen$^{\dagger ,}$\footnote{Corresponding author: Christian.Hansen@cern.ch},
	             N. Gollub$^\dagger$, K. Assamagan$^\ddagger$, T. Ekel\"of$^\dagger$}\\
\vspace{5mm}
$^\dagger$\normalsize {\it Uppsala University - Department of Radiation Sciences}\\
          \normalsize {\it Uppsala, SWEDEN}\\
\vspace{5mm}
$^\ddagger$\normalsize {\it Brookhaven National Laboratory}\\
          \normalsize {\it Upton, NY 11973, USA}\\
\vspace{5mm}
\end{center}
\vspace{5mm}
\begin{abstract}
Charged Higgs boson production via the gluon-bottom quark mode, $gb$~$\to$~$tH^{\pm}$,
followed by its decay into a chargino and a neutralino has been investigated. 
The calculations are based on masses and couplings given
by the Minimal Supersymmetric Standard Model (MSSM)
for a specific choice of MSSM parameters.
The signature of the signal is characterized by three hard leptons, 
a substantial missing transverse energy due to the decay of the neutralino 
and the chargino and three hard jets from the hadronic decay of the top quark.
The possibility of detecting the signal over the Standard Model (SM) and non-SM backgrounds
was studied for a set of $\tan\beta$ and $m_A$. 
The existence of 5-$\sigma$ confidence level regions
for $H^{\pm}$ discovery at integrated luminosities of 100~fb$^{-1}$ and 300~fb$^{-1}$
is demonstrated, which cover also the intermediate region $4$~$\lesssim$~$\tan\beta$~$\lesssim$~10 where
$H^{\pm}$ decays to SM particles cannot be used for $H^{\pm}$ discovery.

\end{abstract}
\end{titlepage} 

\section{Introduction}
\label{intro}

The search for Higgs bosons is at the front-line of present research efforts
in particle physics. While there is a single Higgs boson in the 
Standard Model (SM)~\cite{HIGGS} (the only SM-particle not yet discovered), 
the Minimal Supersymmetric (SUSY) extension of the Standard Model (MSSM) has five of them~\cite{SUSYSEARCH, SUPER}. 
Three neutrals, the $CP$-even $h^0$ and $H^0$ (where $m_h$~$<$~$m_H$),
and the $CP$-odd $A^0$, and two that are charged conjugates of each other, $H^{\pm}$.
The detection of the charged Higgs bosons would unambiguously imply the existence 
of physics beyond the SM, since charged scalar states (like $H^{\pm}$) do not belong
in the SM. 
\begin{figure}[ht]   
\begin{center}
\includegraphics[angle=0, scale= .50]{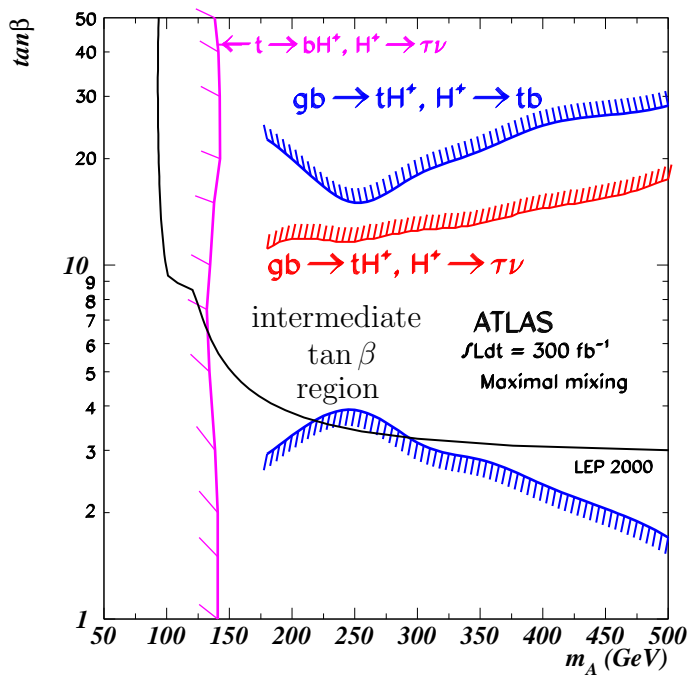}
\put(-186,139){intermediate}
\put(-165,125){$\tan\beta$}
\put(-169,112){region}
\caption{The ATLAS 5-$\sigma$ discovery contour of the charged Higgs \cite{YANN}. 
Below $\sim$160~GeV the processes $t$~$\to$~$bH^{\pm}$,~$H^{\pm}$~$\to$~$\tau\nu$
provides coverage for most $\tan\beta$. Above $\sim$175~GeV the 
$gb$~$\to$~$bH^{\pm}$,~$H^{\pm}$~$\to$~$\tau\nu$ covers the high $\tan\beta$ region
($\tan\beta$~$\gtrsim$~10) while the $H^{\pm}$~$\to$~$tb$ channel covers the
$\tan\beta$~$\lesssim$~4 region. In the intermediate $\tan\beta$ region 
the charged Higgs decays to SM particles are undetectable at the LHC.}
\label{fig:contour}
\end{center}
\end{figure}
The combined LEP collaborations have set lower limits in a model independent
way on the mass of $H^{\pm}$-bosons, $M_{H^{\pm}}$~$>$~78.6 GeV for any value of $\tan\beta$ \cite{LEP}.
At tree level in the MSSM, all Higgs particle masses and couplings are determined by two 
parameters~\cite{JOURNAL}. 
The conventional choice is to use the mass of the $CP$-odd neutral
Higgs, $m_A$, and the ratio of the vacuum expectation values of the Higgs doublets,
$\tan\beta$. For the choice of MSSM parameters considered here, the mass of the $H^{\pm}$
does not differ considerably from that of the $A^0$. 
The discovery potential of $H^{\pm}$ at the LHC has been investigated by both 
ATLAS \cite{YANN} and CMS \cite{DENEGRI} collaborations. It has been established that
for $m_A$ below $\sim$160 GeV and for most values of $\tan\beta$ the charged Higgs can 
be discovered with 95\% confidence level (C.L.) through the process 
$t$~$\to$~$bH^{\pm}$,~$H^{\pm}$~$\to$~$\tau\nu$~\cite{CAVALLI} (see figure~\ref{fig:contour}). 
Above the top quark mass (i.e. $m_{H^{\pm}}$~$\gtrsim$~175~GeV) the charged Higgs is produced via 
the gluon-bottom quark mode, $gb$~$\to$~$tH^{\pm}$.
In this mass region the charged Higgs can be discovered with a 5-$\sigma$ C.L. through
$H^{\pm}$~$\to$~$tb$ for $\tan\beta \lesssim 4$, 
such low $\tan\beta$ values are however already excluded by LEP with the $m_t$ value used for the current 
investigation~\footnote{This exclusion region can however change with the new 
world-averaged $m_t$~=~178.0~$\pm$~4.3~GeV~\cite{TOPMASS}.}, 
and for $\tan\beta \gtrsim 15$~\cite{KETEVI1} and through
$H^{\pm}$~$\to$~$\tau\nu$ for $\tan\beta \gtrsim 10$~\cite{KETEVI2}.
In the intermediate $\tan\beta$ region (4~$\lesssim$~$\tan\beta$~$\lesssim$~10)
charged Higgs decays into SM particles have been found to be
undetectable at the LHC.
This zone of undectectability is for the $H^{\pm}$~$\to$~$tb$ channel partly due to the
$H^{\pm}tb$ Yukawa coupling that is $\sim$~$(m_b \tan\beta + m_t \cot \beta)$ \cite{HTBYUKAWA} and has a minimum
at $\tan\beta = \sqrt{m_t/m_b} \approx 7$.
Also the $H^{\pm}$~$\to$~$\tau\nu$ channel becomes invisible at this region due to decreasing 
expected signal rate for decreasing $\tan\beta$ values~\cite{KETEVI2}.
For heavy charged Higgs ($m_{H^+}$~$>$~$m_t$~+~$m_b$) and for large $\tan\beta$ 
($\tan\beta$~$>$~30 for $m_{H^+}$~$\sim$~250~GeV and $\tan\beta$~$>$~50 for $m_{H^+}$~$\sim$~500~GeV) 
it has been shown that the charged Higgs can be discovered at the LHC through the process 
$gg$~$\to$~$tbH^{\pm}$, $H^{\pm}$~$\to$~$tb$~\cite{NILS}.
\begin{figure}[ht]
\begin{center}
$\begin{array}{c@{\hspace{1in}}c}
\multicolumn{1}{l}{\mbox{\bf (a)}}                 &  \multicolumn{1}{l}{\mbox{\bf (b)}} \\ [-0.53cm]
\includegraphics[angle=0, scale= .40]{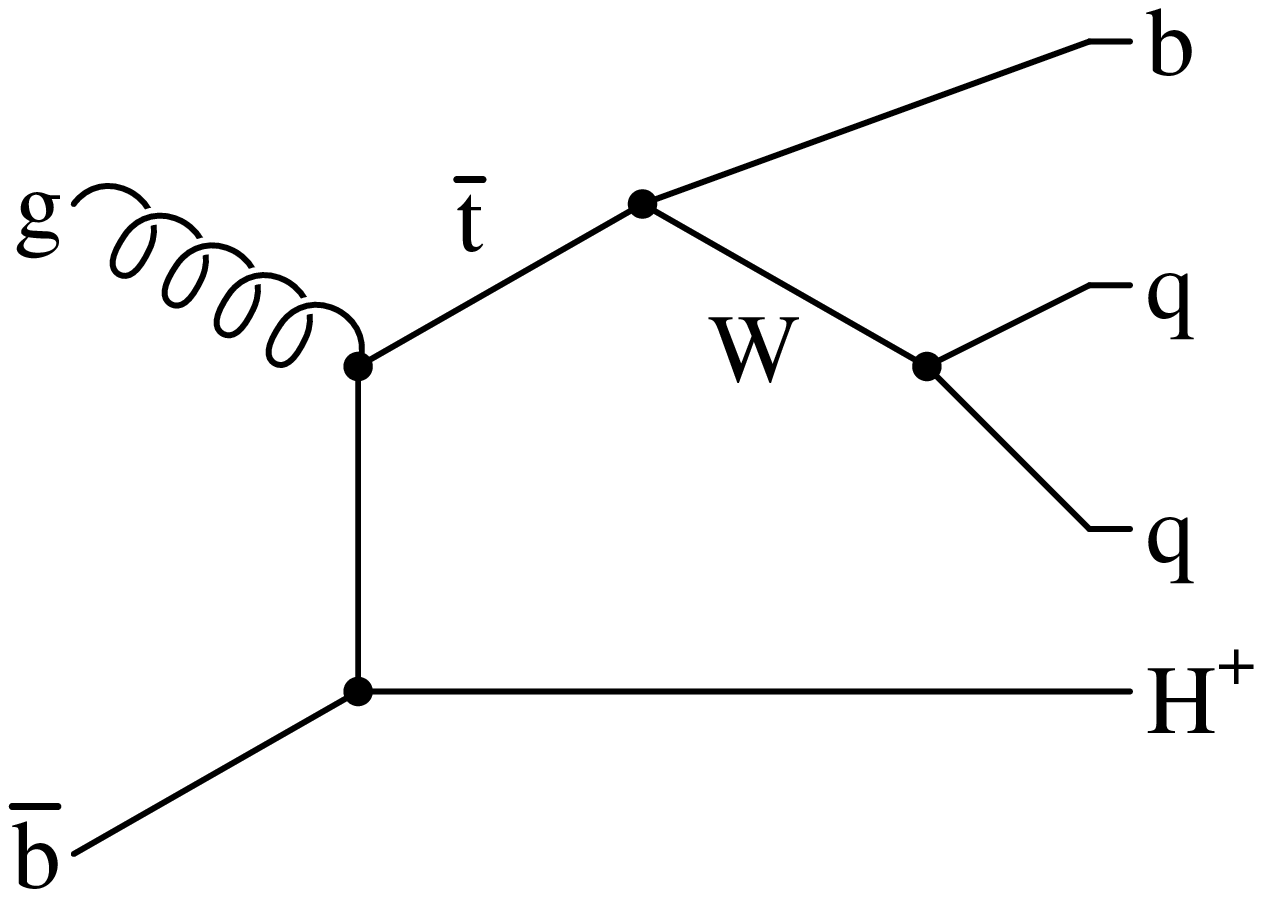}  &  \includegraphics[angle=0, scale= .40]{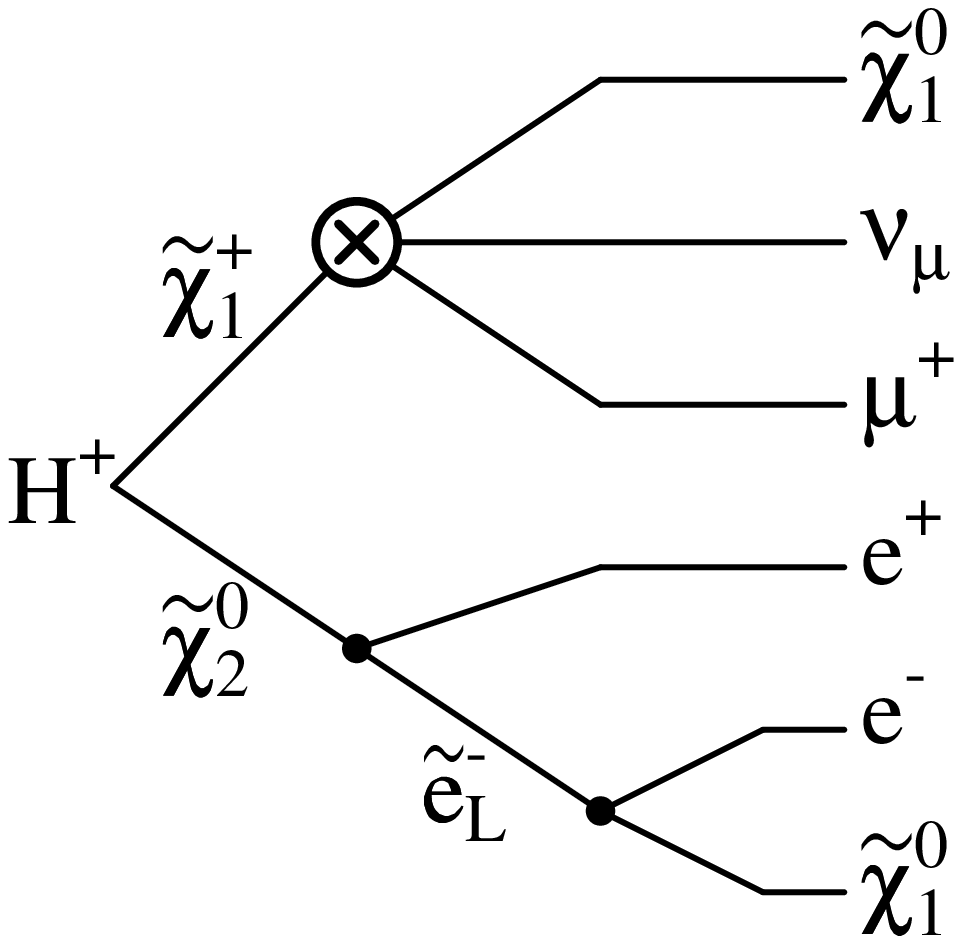} \\ [0.4cm]
\end{array}$
\end{center}
\caption{The Feynman diagram {\bf (a)} shows the production of the $\bar{t}H^+$ final state and the diagram {\bf (b)} shows an 
         example of how $H^{\pm}$ can decay to 3 leptons and several undetectable particles. The $\otimes$ means either
	 $\tilde{\chi}^+_1 \to \tilde{\chi}^0_1 + W^+$ where $W^+ \to \nu_{\mu} + \mu^+$ or 
         $\tilde{\chi}^+_1 \to \tilde{\mu}^+ + \nu_{\mu}$ where $\tilde{\mu}^+ \to \tilde{\chi}^0_1 + \mu^+$ or
	 $\tilde{\chi}^+_1 \to \tilde{\nu}_{\mu} + \mu^+$ where $\tilde{\nu}_{\mu} \to \tilde{\chi}^0_1 + \nu_{\mu}$.} 
\label{fig:diagram}
\end{figure}

Among the new massive sparticles predicted in the MSSM are the charginos, 
$\tilde{\chi}^{\pm}_{1,2}$ and the neutralinos, 
$\tilde{\chi}^0_{1,2,3,4}$, which are mass eigenstate mixtures of the electroweak 
gauginos and the Higgsinos. 
As normally done, the $\tilde{\chi}^0_1$ will be assumed to be the stable Lightest SUSY Particle (LSP),
which is correct unless there is a lighter gravitino.

It is concluded in \cite{CMS} (based on \cite{CMS2}) that searching for $H^{\pm}$ decays into a chargino and a neutralino
can be a viable method for $H^{\pm}$ discovery in parts of the intermediate region $4$~$\lesssim$~$\tan\beta$~$\lesssim$~10. 
The authors show that the processes $g\bar{b}$~$\to$~$\bar{t}H^+$~$+$~$c.c.$ 
with $H^{\pm}$~$\to$~$\tilde{\chi}^{\pm}_{1,2}\tilde{\chi}^0_{1,2,3,4}$ can be distinguished
from the SM and non-SM backgrounds using as signature three hard leptons and 
substantial missing transverse energy from the neutralino and chargino decays and three
hard jets from the hadronic top decay. 
In figure~\ref{fig:diagram}, an example is shown of how the charged Higgs decay
in the chargino neutralino channel
$H^{\pm}$~$\to$~$\tilde{\chi}^{\pm}_{1,2}\tilde{\chi}^0_{1,2,3,4}$~$\to$~$3\ell$~$+$~$N$
can produce three leptons and a number, $N$, of undetectable particles in the final state.
This paper reports on an investigation of the charged Higgs discovery potential for this 
channel in the intermediate $\tan\beta$ region using the ATLAS detector.  
The dominating SM background channels are  $gg \to t\bar{t}$ and $gg \to t\bar{t}Z$.
A third SM background channel, $gg \to t\bar{t}\gamma$, was demonstrated in~\cite{CMS}
to give a negligible contribution.
The dominating non-SM backgrounds are $gg \to t\bar{t}h$, $gg \to \tilde{\chi}\tilde{\chi}$
and $gg \to \tilde{q}, \tilde{g}$.
For this investigation integrated luminosities of 100~fb$^{-1}$ and 300~fb$^{-1}$ are assumed. 
HERWIG~\cite{HERWIG},~\cite{HERWIGSUSY} is used as Monte Carlo physics simulator and
ATLFAST~\cite{ATLFAST} is used to perform a fast simulation of the ATLAS detector\footnote{Both 
packages are used within the ATHENA framework in the ATLAS Software Release 7.0.3.}.

In section~\ref{mssm} the used MSSM parameters and the signal branching ratio (BR) are presented.
In section~\ref{cs} the cross sections for the charged Higgs production and the background channels are discussed. 
Section~\ref{event} shows the expected number of events and in section~\ref{selection} we
describe the analysis. 
Finally the results are summarized and discussed in section~\ref{results}.

\section{MSSM Parameter Point and Signal Branching Ratio}
\label{mssm}

\begin{figure}[ht]
\begin{center}
$\begin{array}{c@{\hspace{0.2in}}c}
\multicolumn{1}{c}{\mbox{\bf (a)}}                 &  \multicolumn{1}{c}{\mbox{\bf (b)}} \\ [-0.1cm]
\includegraphics[angle=0, scale= .37]{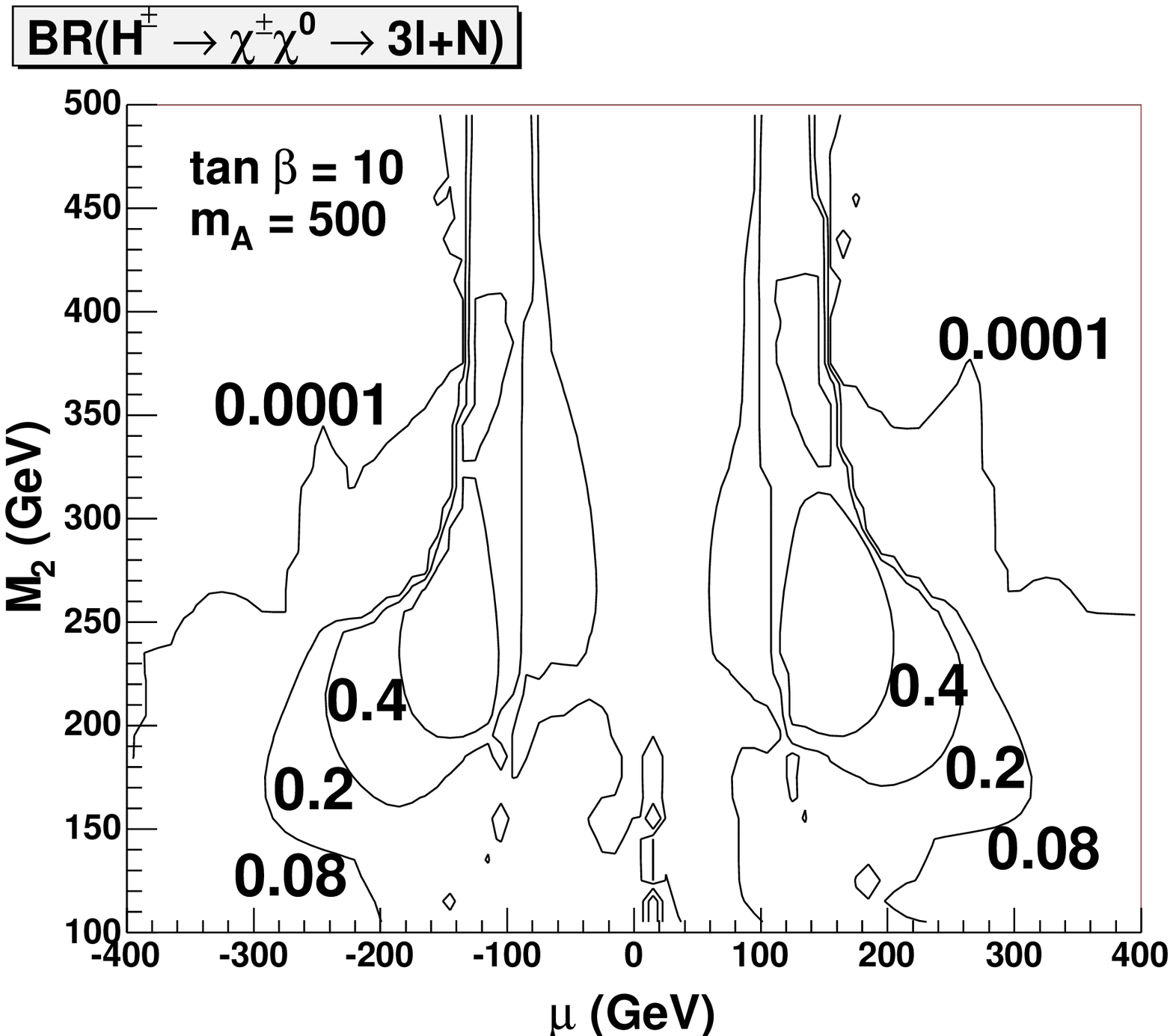}  &  \includegraphics[angle=0, scale= .37]{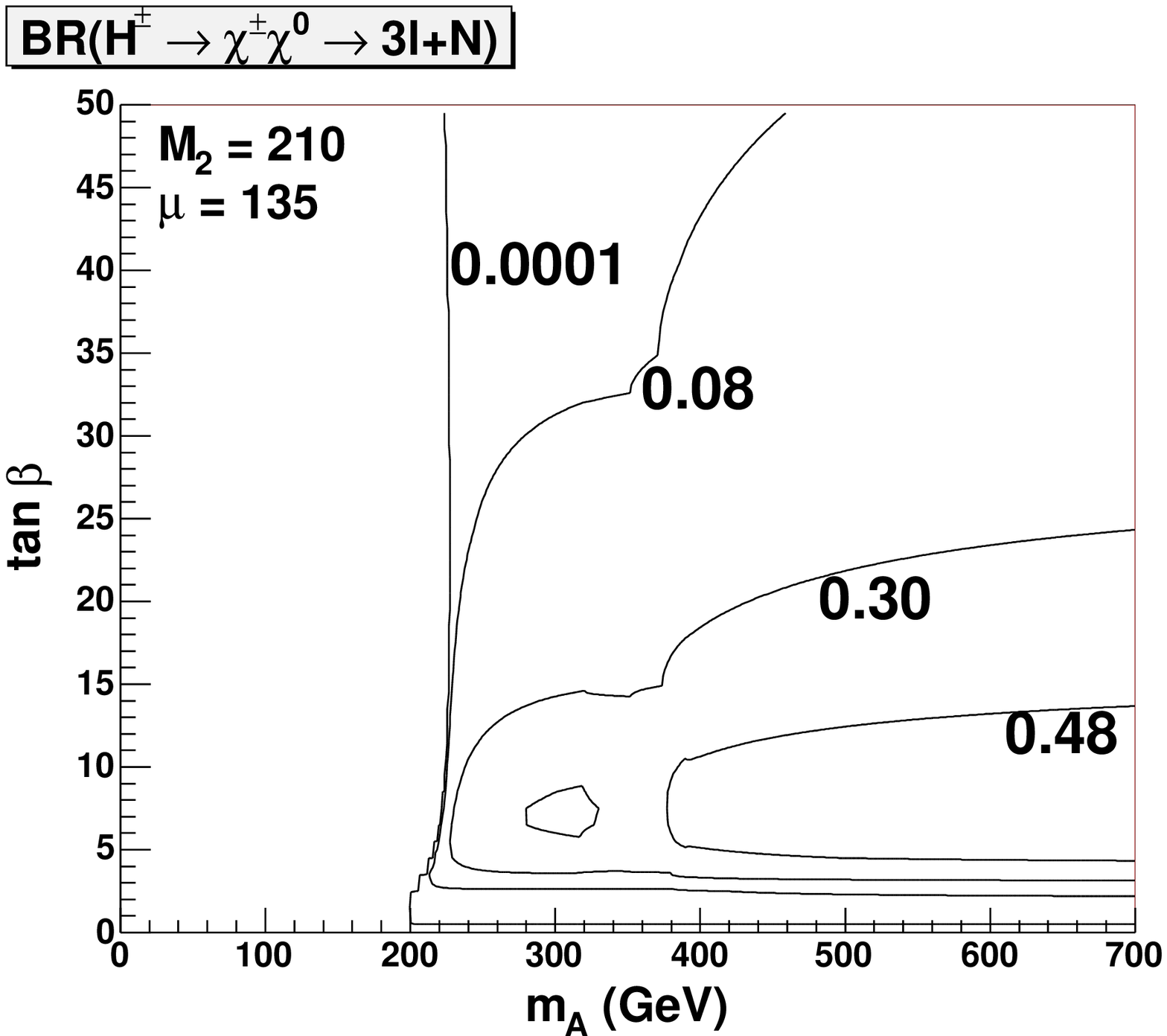} \\ [0.0cm]
\end{array}$
\end{center}
\caption{The contours of constant branching ratio of $H^{\pm}$~$\to$~$\tilde{\chi}^{\pm}_{1,2}\tilde{\chi}^0_{1,2,3,4}$~$\to$~$3\ell+N$, 
         where $N$ represents undetectable final state particles and $\ell$~=~$e^{\pm}$ or $\mu^{\pm}$,
         are shown in {\bf(a)} for $\tan\beta~=~10$ and $m_A~=~500$~GeV in the $M_2$ vs. $\mu$ plane 
	 and in {\bf (b)} for $M_2~=~210$~GeV and $\mu~=~135$~GeV in the $\tan\beta$ vs. $m_A$ plane.
         Both in {\bf(a)} and {\bf (b)} $m_{\tilde{\ell}_R}$~=~110~GeV, $m_{\tilde{\tau}_R}$~=~210~GeV,
	 $m_{\tilde{g}}$~=~800~GeV, $m_{\tilde{q}}$~=~1~TeV and $m_t$~=~175~GeV and the trilinear coupling terms 
	 are set to zero.}
\label{fig:br}
\end{figure}
For the analysis shown in this paper a point in the MSSM parameter space is chosen
so that the branching ratio for Higgs decays into a chargino and neutralino are maximized.
The same point was used for the CMS-analysis in~\cite{CMS}.
Many independent input MSSM parameters are involved in the calculation of the 
$H^{\pm}$~$\to$~chargino-neutralino rate.
At tree-level, two parameters, $m_A$ and $\tan\beta$, completely specify the Higgs masses and
couplings to SM particles~\cite{JOURNAL}. 
For the charginos and neutralinos (the ``inos'') the tree-level masses and couplings to the charged Higgs bosons
are determined by the parameters $m_A$, $\tan\beta$, $M_2$ and $\mu$. $M_1$ is assumed to be determined 
from $M_2$ via gaugino unification; $M_1$~=~$\frac{5}{3}\tan^2\theta_WM_2$. 
\begin{figure}[ht]
\begin{center}
$\begin{array}{c@{\hspace{0.2in}}c}
\multicolumn{1}{c}{\mbox{\bf (a)}}                                &  \multicolumn{1}{c}{\mbox{\bf (b)}} \\ [-0.1cm]
\includegraphics[angle=0, scale= .30]{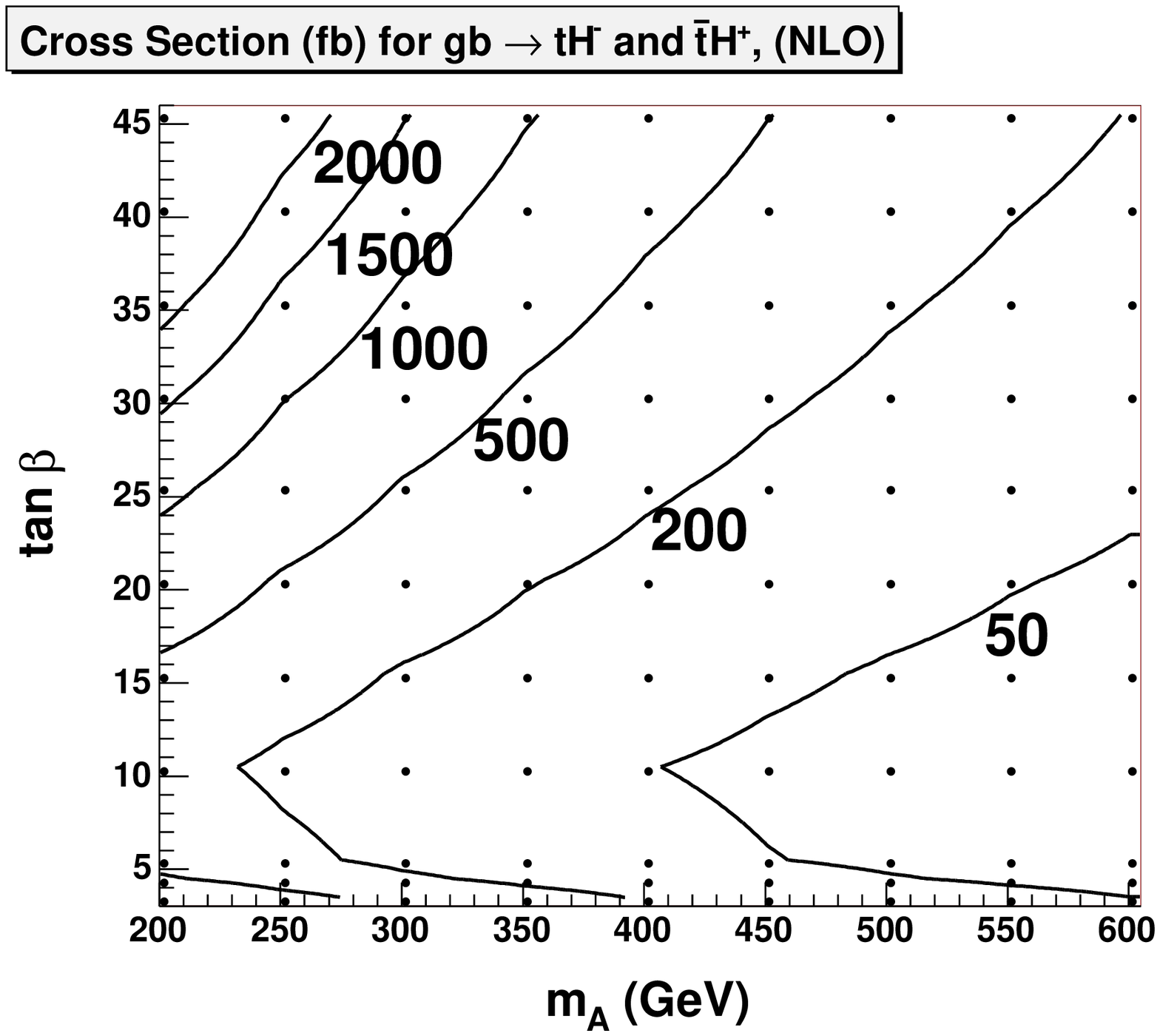}  &  \includegraphics[angle=0, scale= .30]{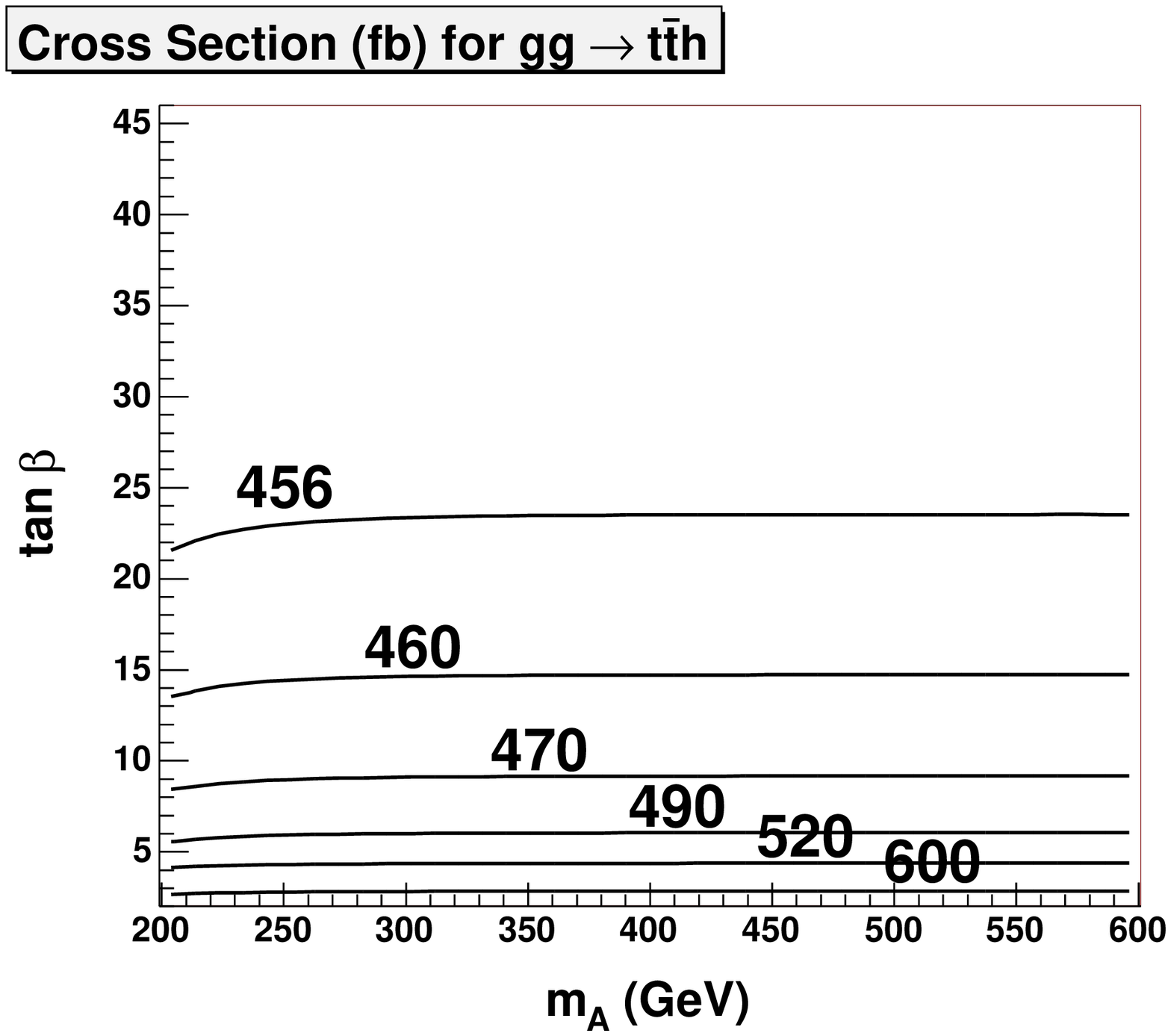} \\ [0.0cm]
\multicolumn{1}{c}{\mbox{\bf (c)}}                                &  \multicolumn{1}{c}{\mbox{\bf (d)}} \\ [-0.1cm]
\includegraphics[angle=0, scale= .30]{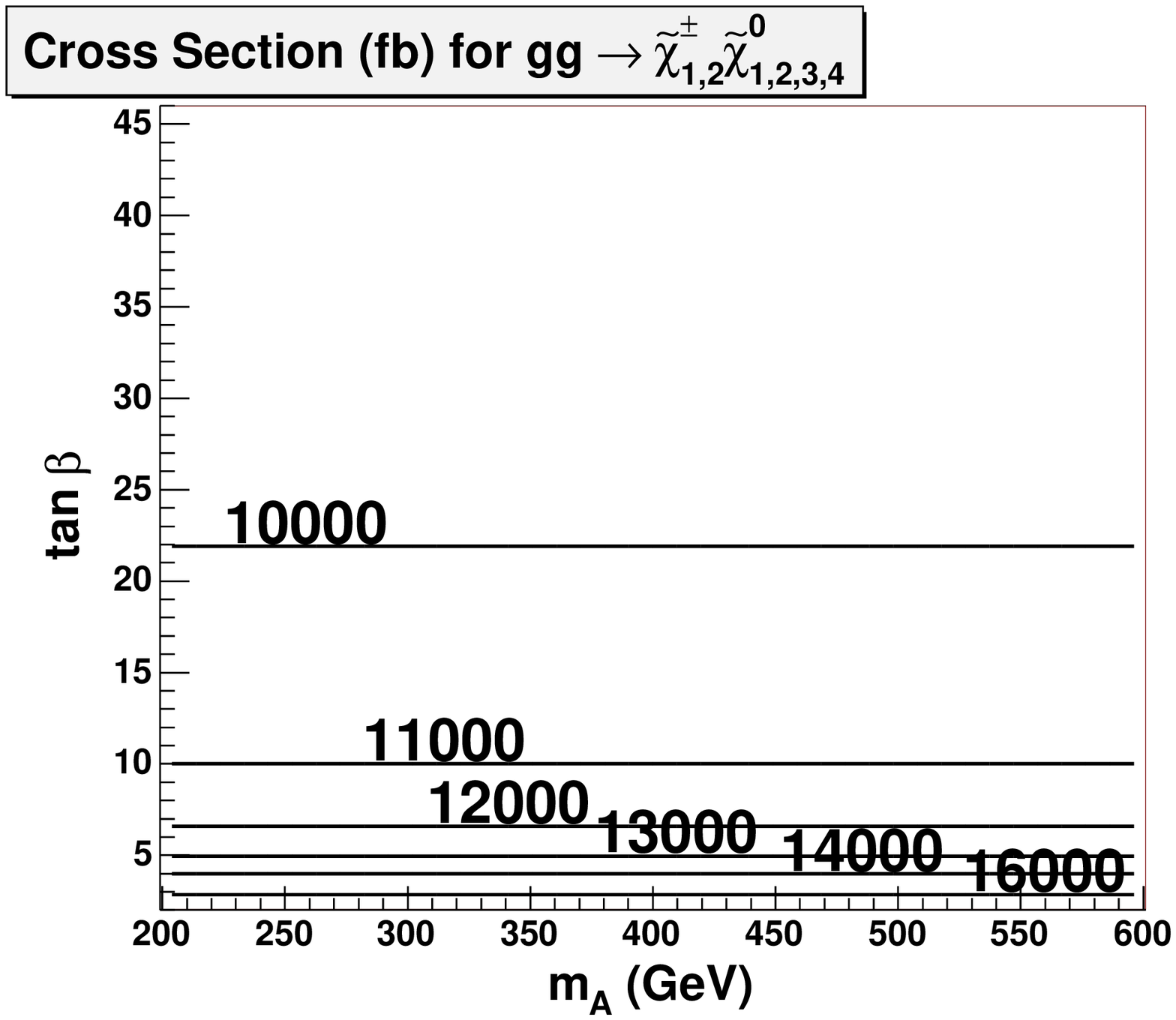}   &  \includegraphics[angle=0, scale= .30]{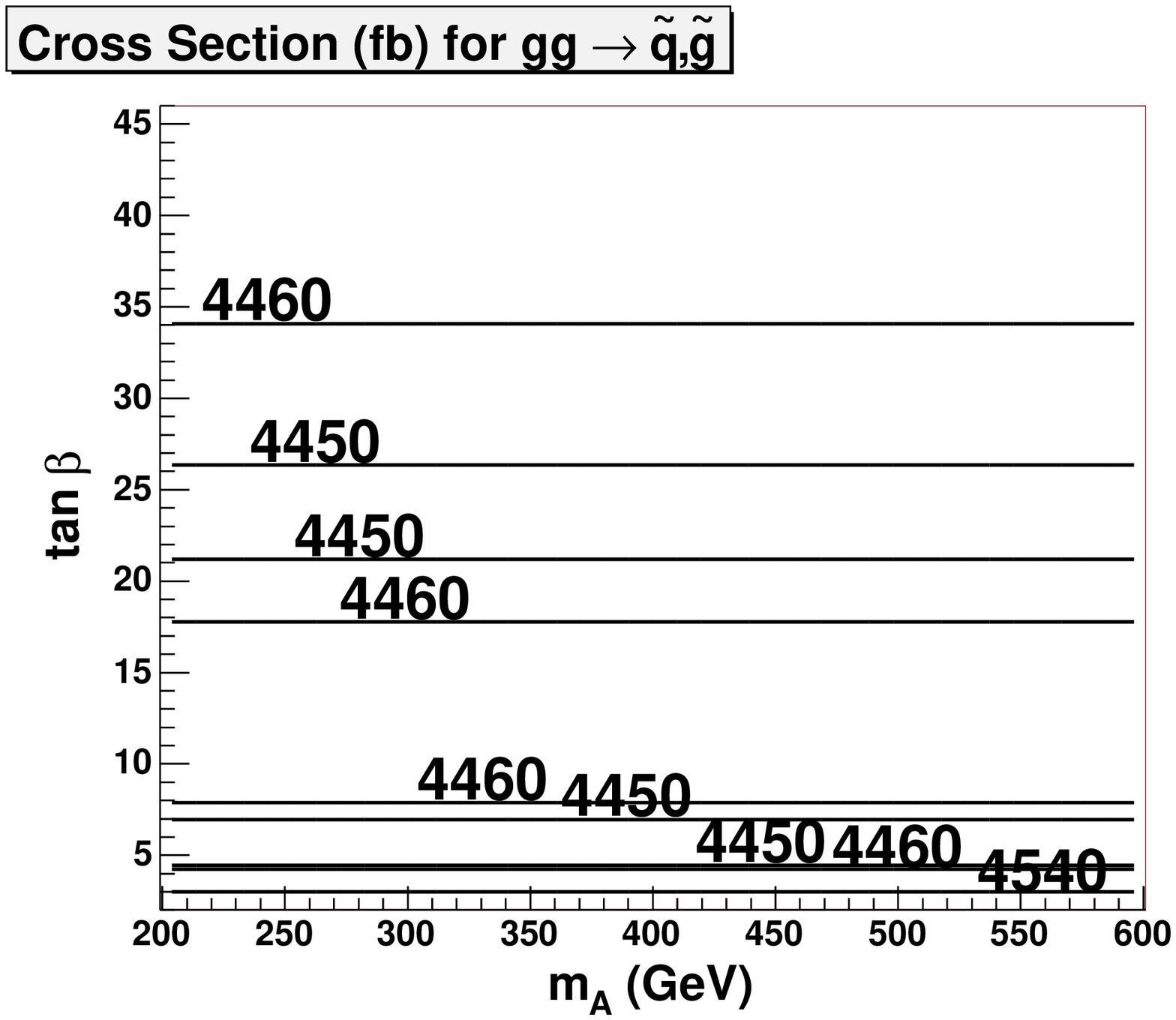} \\ [0.0cm]
\end{array}$
\end{center}
\caption{The contours in {\bf (a)} shows the NLO cross section for the signal ($gb \to tH^+, \bar{t}H^-$). 
         The cross sections for the ($\tan\beta, m_A$) values indicated with dots are taken from~\cite{TILMAN} 
	 and in between the cross section is calculated with a linear interpolation. 
         In {\bf (b)}, {\bf (c)} and  {\bf (d)} the contours for the LO cross section 
	 (obtained from HERWIG) 
	 for the SUSY background channel $gg \to t\bar{t}h$, $gg \to \tilde{\chi}\tilde{\chi}$
         and $gg \to \tilde{q}, \tilde{g}$ respectively are shown.}
\label{fig:cs}
\end{figure}
The branching ratio of 
ino decays to leptons depend on the trilinear coupling, $A_{\tau}$, 
which is here put to zero, and on the left- and right-handed soft slepton masses. 
It is assumed that for all three generations 
$m_{\tilde{\ell}_R}$~=~$m_{\tilde{\ell}_L}$ and that the two first generations are degenerate in mass,
i.e. $m_{\tilde{e}_{L, R}}$~=~$m_{\tilde{\mu}_{L, R}}$~$=$~$m_{\tilde{\ell}_R}$. 
It was chosen to maximize the ino decays to leptons by setting $m_{\tilde{\ell}_R}$ to the lowest value 
allowed by the LEP results which is $m_{\tilde{\ell}_R}$~$\sim$~110~GeV \cite{CMS}.  
The third generation slepton masses, $m_{\tilde{\tau}_{L, R}}$, are assumed to be
100~GeV above the selectron and smuon masses. This enhances the ino decays to leptons since the branching ratios
of ino~$\to$~stau are then less than to the first slepton generations.
A heavy gluino mass, $m_{\tilde{g}}$~=~800~GeV, is assigned and all squark masses are set to 1~TeV.
These values are used as input to ISAJET7.64~\cite{ISAJET769} to calculate the total branching ratio of 
$H^{\pm}$~$\to$~$\tilde{\chi}^{\pm}_{1,2}\tilde{\chi}^0_{1,2,3,4}$~$\to$~$3\ell$~$+$~$N$
in the $M_2$-$\mu$ parameter space setting $\tan\beta$~=~10 and $m_A$~=~500~GeV. 
According to the results shown in figure~\ref{fig:br}a~\footnote{ISAWIG 1.200 
(an ISAJET-HERWIG interface) was used for these calculations.}
the branching ratio of charged Higgs decay to
three leptons is higher than 0.4 by choosing $M_2$~=~210~GeV and $\mu$~=~135~GeV.
The MSSM parameter point chosen is thus:
\begin{itemize}
\item $M_2$~=~210~GeV, $\mu$~=~135~GeV, $m_{\tilde{\ell}_R}$~=~110~GeV,  $m_{\tilde{\tau}_R}$~=~210~GeV, 
  $m_{\tilde{g}}$~=~800~GeV, $m_{\tilde{q}}$~=~1~TeV.
\end{itemize}
With these values as input ISAJET7.64 was used to calculate the total branching ratio of 
$H^{\pm}$~$\to$~$\tilde{\chi}^{\pm}_{1,2}\tilde{\chi}^0_{1,2,3,4}$~$\to$~$3\ell$~$+$~$N$
in the $\tan\beta$-$m_A$ space. The results are shown in figure~\ref{fig:br}b.

\section{Cross Sections}
\label{cs}

The next-to-leading order (NLO) cross section for the signal, $gb$~$\to$~$tH^{\pm}$, has been calculated in~\cite{TILMAN} 
for the $\tan\beta$ and $m_A$ values indicated with dots in 
figure~\ref{fig:cs}a~\footnote{In \cite{TILMAN} the 
cross sections are given for ($\tan\beta$, $m_H$) values. 
The change in the cross section when putting $m_H$~=~$m_A$ is negligible.}.
The contours of constant cross section in this plot have been obtained by linear interpolation between the points.
The NLO cross section for the SM background channel $t\bar{t}$ is 737~pb~\cite{TOP}.
For the $t\bar{t}Z$ and the non-SM background channels 
only leading order (LO) cross section are used.
However, the $t\bar{t}$ background is by far the largest of all the backgrounds
and the NLO corrections to the other background channels would be small with respect to the $t\bar{t}$ alone.
The LO cross section for the SM background channel $t\bar{t}Z$ as obtained with HERWIG is 439~fb. 
The LO cross section for the SUSY background channels depend on $\tan\beta$ and weakly on $m_A$ and are calculated using
HERWIG with its SUSY extension~\cite{HERWIGSUSY}. The obtained result is shown for the 
SUSY background channels $gg \to t\bar{t}h$, $gg \to \tilde{\chi}\tilde{\chi}$
and $gg \to \tilde{q}, \tilde{g}$ in figure~\ref{fig:cs}b, c and d respectively.

\section{Event Production}
\label{event}

Table~\ref{tab:ev} shows for each process the number of simulated events generated by HERWIG and processed
with ATLFAST.  
The signal and SUSY background events were produced for the $\tan\beta$ and $m_A$ 
values represented by the 99 dots in figure~\ref{fig:cs}a.
The SM backgrounds obviously do not depend on any MSSM parameter. 
\begin{table}[htbp]
\begin{center}
\begin{tabular}{l|l|l}
\hline
\hline
         & Process                                                                                                         & Number Events Produced \\
\hline
\hline
Signal   & $gb$~$\to$~$tH^{\pm}$, $H^{\pm}$~$\to$~$\tilde{\chi}^{\pm}_{1,2}\tilde{\chi}^0_{1,2,3,4}$~$\to$~$3\ell$~$+$~$N$ & $\sim$4$\cdot 10^5$ for each ($\tan\beta$, $m_A$) \\
         & and $t$ $\to$ $bqq$                                                                                             & dot in figure~\ref{fig:cs}a \\
\hline  
SM Bkg   & $gg$~$\to$~$t\bar{t}$                                                                                           & 10$^8$ \\
\hline  
         & $gg$~$\to$~$t\bar{t}Z$                                                                                          & 2$\cdot 10^7$ \\
\hline
SUSY Bkg & $gg$~$\to$~$t\bar{t}h$                                                                                          & 10$^7$ for each ($\tan\beta$, $m_A$) dot \\
\hline  
         & $gg$~$\to$~$\tilde{\chi}\tilde{\chi}$                                                                           & 10$^7$ for each ($\tan\beta$, $m_A$) dot \\
\hline
         &  $gg$~$\to$~$\tilde{q}, \tilde{g}$                                                                              & 10$^6$ for each ($\tan\beta$, $m_A$) dot \\ 
\hline
\hline
\end{tabular}
\caption{In the rightmost column the number events produced for each process,
         ensuring a negligible statistical uncertainty on the number of expected event after all cuts, are shown.
         The signal and SUSY background events were produced for the different $\tan\beta$ and $m_A$ values 
         represented by the dots in figure~\ref{fig:cs}a. The SM background 
	 processes do not depend on MSSM parameters.}
\label{tab:ev}
\end{center}
\end{table}

\section{Event Selection}
\label{selection}

Four different cuts were optimized and used to enhance the signal over background ratio 
in the generated event sample. 
The first cut excludes events without three isolated leptons. 
This {\it Three Lepton Cut} requires:
\begin{itemize}
\item Exactly three isolated leptons ($\ell$~=~$e$ or $\mu$) with  $|\eta|$~$<$~2.4,
      with $p_T$~$>$~7~GeV and at least one of which with $p_T$~$>$~20~GeV.
\end{itemize}
An electron (muon) is considered as an isolated lepton if it has 
$p_T$~$>$~5~GeV~(6~GeV), if it is within $|\eta|$~$<$~2.5, if 
the sum of energy deposited in a hollow cone of
0.1~$<$~$\Delta R$~$<$~0.4 (where $\Delta R$~=~$\sqrt{(\Delta\varphi)^2+(\Delta\eta)^2}$) 
around the lepton track is less than 10~GeV and 
if no other isolated lepton nor jet is within the cone $\Delta R$~=~0.4 around it~\cite{ATLFAST}.
In order to trigger on the signal events the detection of high-$p_T$ leptons
are required. The first cut mentioned above is chosen such that it meet the
requirements of the ATLAS trigger system which has rather low $p_T$ thresholds 
on electrons and muons~\cite{ATLASTRIGGER}.
\begin{figure}[ht]
\begin{center}
$\begin{array}{c@{\hspace{0.1in}}c}
\multicolumn{1}{l}{\mbox{\bf (a)}}                 &  \multicolumn{1}{l}{\mbox{\bf (b)}} \\ [0.10cm]
\includegraphics[angle=0, scale= .4]{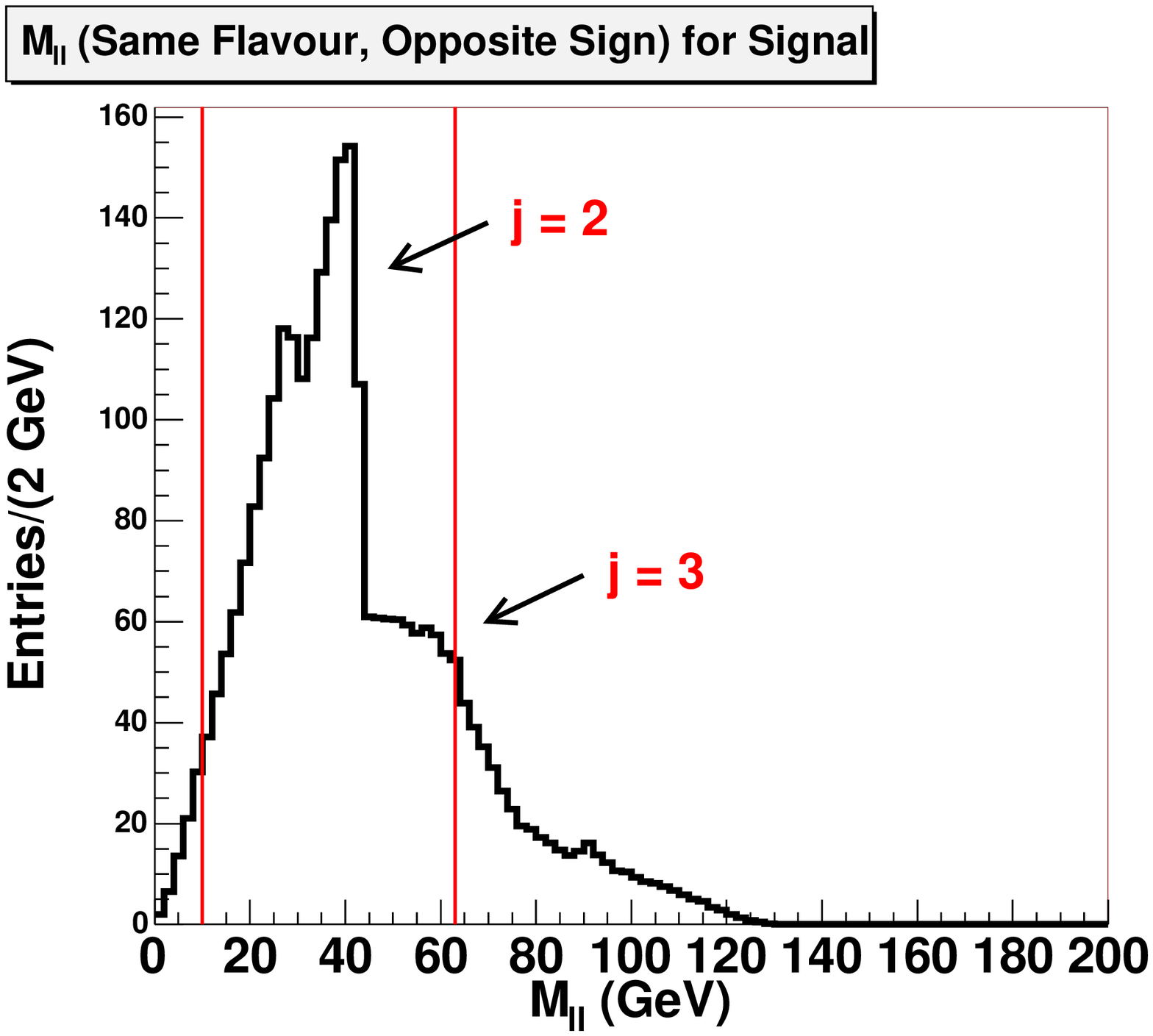}  &  \includegraphics[angle=0, scale= .4]{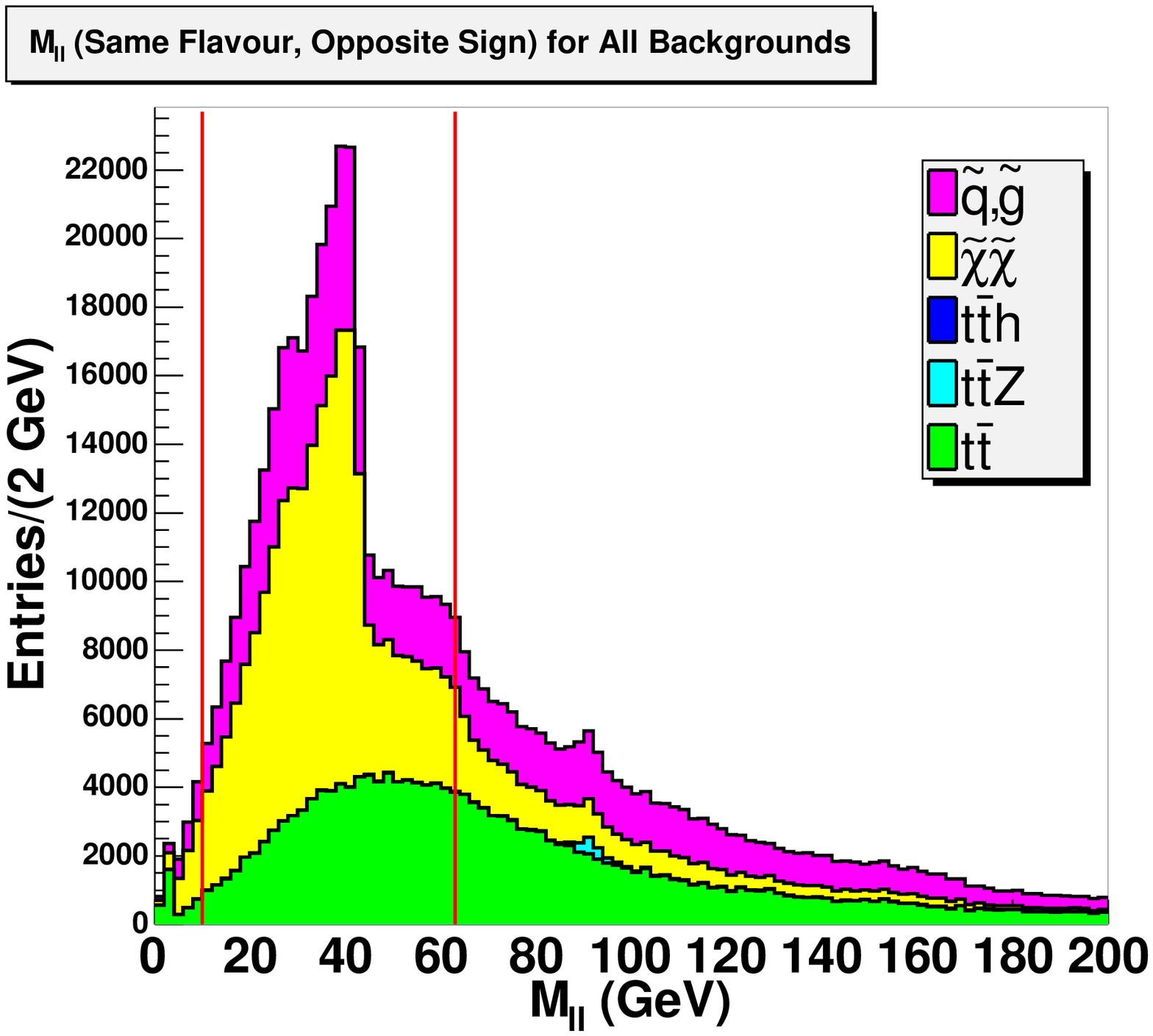} \\ [0.2cm]
\end{array}$
\end{center}
\caption{Invariant mass, $M_{\ell\ell}$, of two isolated leptons with the same flavor and opposite charge is histogrammed 
         for {\bf (a)} the signal and for {\bf (b)} the background channels. 
         For the SUSY channels $\tan\beta$~=~10 and $m_A$~=~350~GeV. 
         The histograms are normalised to the expected event rate for an integrated luminosity of 300~fb$^{-1}$.}
\label{fig:2l}
\end{figure}

The second cut is based on the fact that for the signal two of the three isolated leptons
come from the decay of a neutralino, $\tilde{\chi}^0_j$. The neutralino can 
decay to two same flavor, opposite charge leptons plus some undetectable particles. 
The invariant mass of the lepton pair, $M_{\ell\ell}$,
is kinematically constrained. For example when $\tilde{\chi}^0_1$ is the only undetectable
particle from $\tilde{\chi}^0_j$ (as in figure~\ref{fig:diagram}b) we have~\cite{SUSYMEAS}
\begin{equation}
M_{\ell\ell_{max}} = \sqrt{\left(m^2_{\tilde{\chi}^0_j} - m^2_{\tilde{\ell}}\right)
    \left(m^2_{\tilde{\ell}} - m^2_{\tilde{\chi}^0_1}\right) / m^2_{\tilde{\ell}}}  \ \ ,
\label{mll}
\end{equation}
where $\tilde{\ell}$ is the slepton involved in the decay chain $\tilde{\chi}^0_j \to \ell\tilde{\ell} \to 2\ell + \tilde{\chi}^0_1$.
For $\tan\beta$~=~10 and $m_A$~=~350~GeV we have $m_{\tilde{\chi}^0_1}$ = 77.6~GeV,
$m_{\tilde{\chi}^0_2}$ = 131.4~GeV and $m_{\tilde{\chi}^0_3}$ = 145.9~GeV for the MSSM point given in section~\ref{mssm}
~\cite{ISAJET769}. This gives $M_{\ell\ell_{max}}$~=~50.9~GeV~(68.0~GeV) for $j = 2$ ($j = 3$) 
which can be approximately seen by the first (second) edge in figure~\ref{fig:2l}a. The
explanation for the tail above $M_{ll}$~$>$~63~GeV is limited invariant mass resolution in this region.
For $\tan\beta$ = 10 and $m_A$ = 350~GeV $m_{H^+}$~$<$~$m_{\tilde{\chi}^0_4}$ + $m_{\tilde{\chi}^+_1}$,
for the MSSM point given in section~\ref{mssm}, which makes a third edge (for j = 4) kinematically impossible.  
The edge corresponding to $j = 3$ is not very dependent on $\tan\beta$ or $m_A$ and 
corresponds approximately to the maximum value for $M_{\ell\ell}$ (see figure~\ref{fig:2l}a and b).
A minimum value of $M_{\ell\ell}$~=~10~GeV and a maximum value of 63~GeV is allowed.
The {\it Two Lepton Cut} requires:
\begin{itemize}
\item Out of the three isolated leptons at least one pair is found with two lepton of same flavour, opposite sign 
      and with invariant mass in the range 10~GeV~$< M_{\ell\ell} < 63$~GeV.
\end{itemize}

Apart from requirements on the isolated leptons, it is also possible to impose conditions on the 
hadronic decay products of the top quark ($t \to b+W$) that is produced together with the charged Higgs boson.
The {\it Top Cut} requirements are:
\begin{itemize}
\item Events must have at least three jets, each with $p_T > 20$~GeV in $|\eta| < 4.5$.
\item Among these, the three jets most likely to come from the top quark are selected by 
      minimizing $|m_{jjj} - m_t|$, where $m_{jjj}$ is the invariant mass of the three-jet system.
      It is required that $|m_{jjj} - m_t| < 35$~GeV.
\item Among these three top jets, the two jets most likely to come from the $W$ boson is selected by
      minimizing  $|m_{jj} - m_W|$, where $m_{jj}$ is the invariant mass of the two-jet system.
      It is required that $|m_{jj} - m_W| < 15$~GeV.
\end{itemize}
\begin{figure}[tp]
\begin{center}
$\begin{array}{c@{\hspace{0.1in}}c}
\multicolumn{1}{l}{\mbox{\bf (a)}}                      &  \multicolumn{1}{l}{\mbox{\bf (b)}} \\ [0.10cm]
\includegraphics[angle=0, scale= .4]{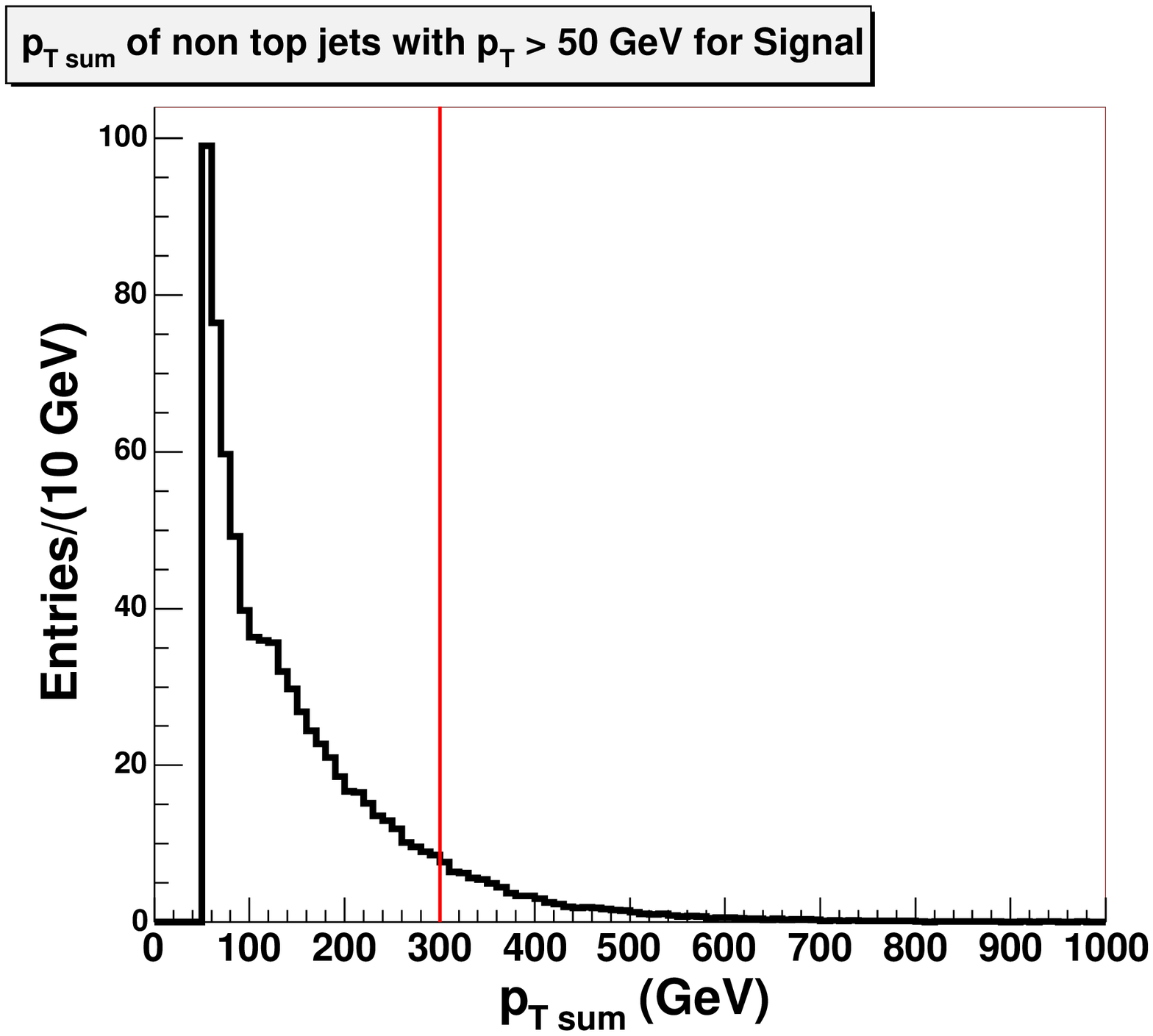}  &  \includegraphics[angle=0, scale= .4]{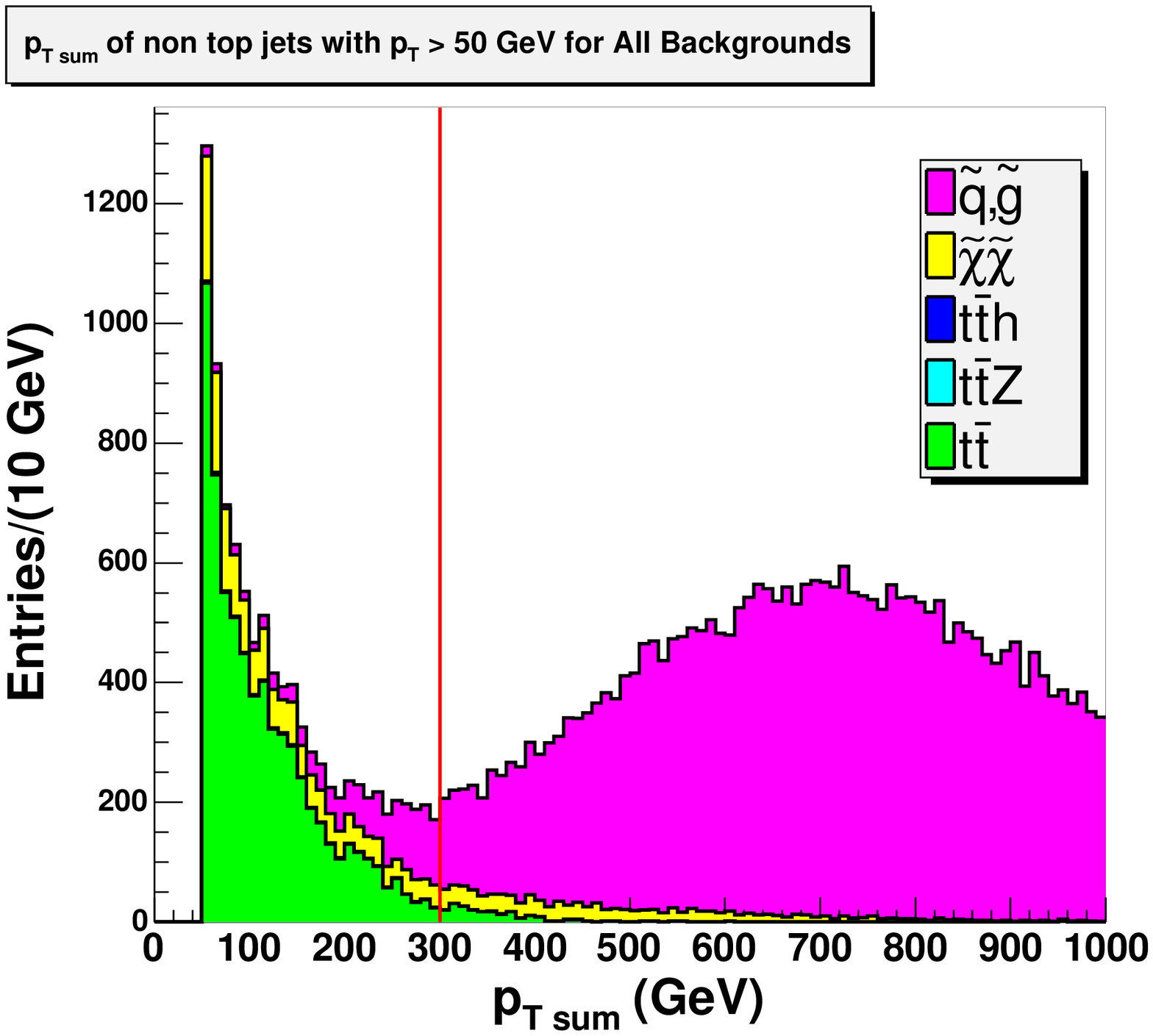} \\ [0.2cm]
\end{array}$
\end{center}
\caption{The sum of $p_T$ for jets that have $p_T$~$>$~50~GeV and $|\eta|$~$<$~4.5, but were not assigned to one of the three top jets for
         {\bf (a)} the signal and for {\bf (b)} the background channels. For the SUSY channels $\tan\beta$~=~10 and $m_A$~=~350~GeV.
         The histograms are normalised to the expected event rate for an integrated luminosity of 300~fb$^{-1}$.}
\label{fig:ptsum}
\end{figure}
The requirement that the third jet (assigning the two first to the W decay) be tagged as a $b$-jet was also tested.
The $b$-tagging of jets is simulated in ATLFAST and even though a $b$-tagging efficiency as high as 0.6 was assumed
the {\it $b$-Tag Cut} did not increase the significance.

The remaining jets that pass the cuts of $p_T > 20$~GeV and $|\eta|$~$<$~4.5 but which where not assigned to the top quark
are used for the fourth cut. 
The $\tilde{q}, \tilde{g}$ background contains many hard jets as shown in figure~\ref{fig:ptsum}. 
In view of this, the {\it Jet Cut} requires:
\begin{itemize}
\item The scalar sum of the $p_T$ of all jets with $p_T > 50$~GeV, excluding jets assigned to the top quark, is not allowed to
      exceed 300~GeV.
\end{itemize}

\section{Results}
\label{results}
\begin{figure}[tp]   
\begin{center}
\includegraphics[angle=0, scale= .50]{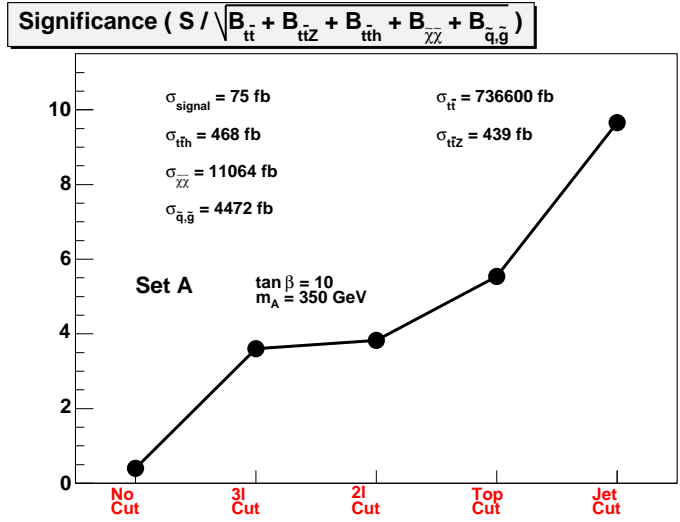}
\caption{The significance ($S/\sqrt{B}$) after each cut is shown for
         $\tan\beta = 10$ and $m_A$~=~350~GeV. 
         The other MSSM parameter values are $M_2$~=~210~GeV, $\mu$~=~135~GeV, $m_{\tilde{\ell}_R}$~=~110~GeV,  
         $m_{\tilde{\tau}_R}$~=~210~GeV, $m_{\tilde{g}}$~=~800~GeV, $m_{\tilde{q}}$~=~1~TeV.
         An integrated luminosity of~300~fb$^{-1}$
         is assumed. The cross sections ($\sigma$) are taken from figure~\ref{fig:cs}.} 
\label{fig:significance}
\end{center}
\end{figure}
For three years of LHC operation at high luminosity (10$^{34}$~cm$^{-2}$s$^{-1}$)  
the integrated luminosity is about L~=~300~fb$^{-1}$.
The total number of events collected during this period for a certain process is $N_{tot}$~=~BR$\cdot\sigma\cdot$L, 
where the branching ratio for the signal is obtained from figure~\ref{fig:br} 
and the production cross sections, $\sigma$, for the signal and the backgrounds from figure~\ref{fig:cs}.
\begin{figure}[tp]   
\begin{center}
\includegraphics[angle=0, scale= .60]{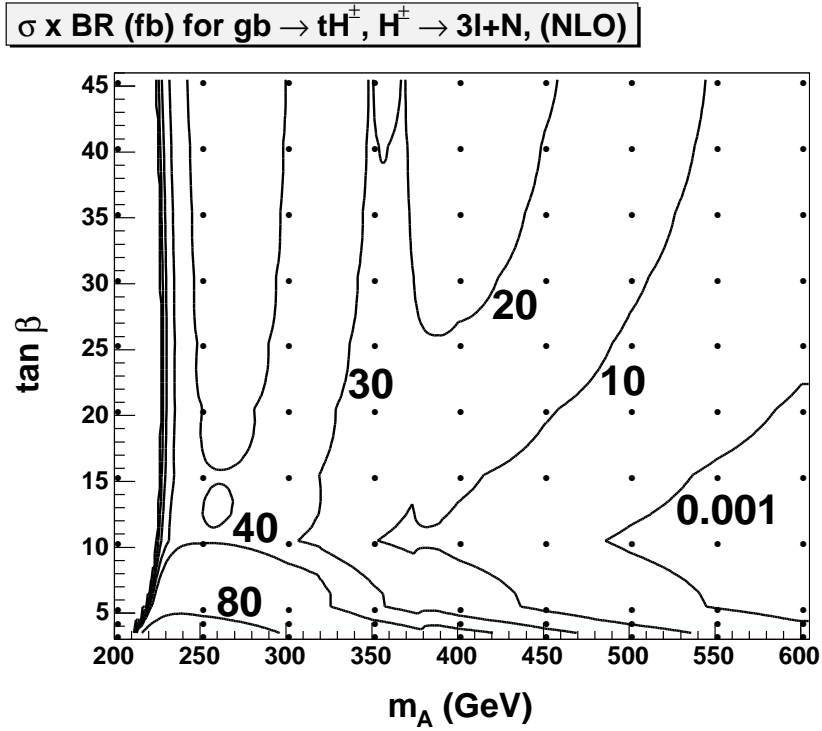}
\caption{The expected rate ($\sigma \times$BR) in fb for $gb \to tH^{\pm}$ with $H^{\pm} \to 3l + N$.} 
\label{fig:csxbr}
\end{center}
\end{figure}
The product of the cross section and the branching ratio for $gb \to tH^{\pm}$ with $H^{\pm} \to 3l + N$ can be seen in figure~\ref{fig:csxbr}.
To achieve the expected signal rate a factor of $2/3$, for the hadronical top decay, has to be multiplied with the values in figure~\ref{fig:csxbr}. 
For the generation of the backgrounds the decay modes are not restricted and hence BR~=~1 for all background processes.
If the efficiency for a certain selection is $\varepsilon_{sel}$, then 
the number $N$ of events surviving this selection is $N = \varepsilon_{sel} N_{tot}$. 
\begin{table}[htbp]
\begin{center}
\begin{scriptsize}
\begin{tabular}{l|l|l|l|l|l|l|c}
\hline
\hline
        & $N_{signal}$   &  $N_{t\bar{t}}$   & $N_{t\bar{t}Z}$   & $N_{t\bar{t}h}$ & $N_{\bar{\chi}\bar{\chi}}$ & $N_{\bar{q},\bar{g}}$   &$\frac{S}{\sqrt{B}}$   \\
\hline
\hline
No Cut  &  6068.7        & 2.2$\cdot 10^{8}$ & 131738          & 140456          & 3.3$\cdot 10^{6}$          & 1.3$\cdot 10^{6}$    & 0.4   \\
\hline  
3l Cut  & 3018.3$\pm$3.3 & 250938$\pm$744 & 2554.0$\pm$4.1   & 11.6$\pm$0.3    & 241851$\pm$273           & 207915$\pm$343     & 3.6 \\ 
\hline  
2l Cut  & 2246.1$\pm$3.2 & 87201$\pm$439 & 129.1$\pm$0.9   & 6.4$\pm$0.2     & 174724$\pm$234           & 82329$\pm$228  & 3.8 \\ 
\hline
Top Cut & 1327.6$\pm$2.8 & 12220$\pm$164 & 55.2$\pm$0.6   & 4.4$\pm$0.2     & 4134$\pm$37            & 41074$\pm$163    & 5.5 \\ 
\hline  
Jet Cut & 1239.7$\pm$2.7 & 11995$\pm$163 & 47.4$\pm$0.5  & 3.2$\pm$0.2     & 3092$\pm$32            & 1331$\pm$30      & 9.7 \\ 
\hline
\hline
\end{tabular}
\caption{Number of events for signal and backgrounds after the successive introduction of the four cuts,
         for $\tan\beta = 10$, $m_A = 350$~GeV and 
         for an integrated luminosity of L~=~300~fb$^{-1}$.
         The errors quoted are derived from the Monte Carlo statistical error.
         In the last column the significance is shown.}
\label{tab:res}
\end{scriptsize}
\end{center}
\end{table}
The resulting numbers $N$ are given in table~\ref{tab:res} for all four cuts introduced in succession 
for $\tan\beta = 10$, $m_A = 350$~GeV, the values
of the other MSSM parameters as specified in section~\ref{mssm} and L~=~300~fb$^{-1}$.
The significance is calculated as
\begin{equation}
\mathcal{S} = \frac{N_{signal}}{\sqrt{\sum_{i} N_{bkg_i}}} \ ,
\label{signif}
\end{equation}
where the sum is taken over all background channels (see table~\ref{tab:ev}).
In the analysis the cuts have been chosen so as to maximize the final significance,
i.e. the significance after the Jet Cut for $\tan\beta = 10$ and $m_A = 350$~GeV.
The progression of the signal-to-background rejection when imposing in sequence the four different cuts
is visualized in figure~\ref{fig:significance} where the significance
after each cut is plotted.

\begin{figure}[ht]
\begin{center}
$\begin{array}{c@{\hspace{0.2in}}c}
\multicolumn{1}{c}{\mbox{\bf (a)}}                                &  \multicolumn{1}{c}{\mbox{\bf (b)}} \\ [-0.1cm]
\includegraphics[angle=0, scale= .30]{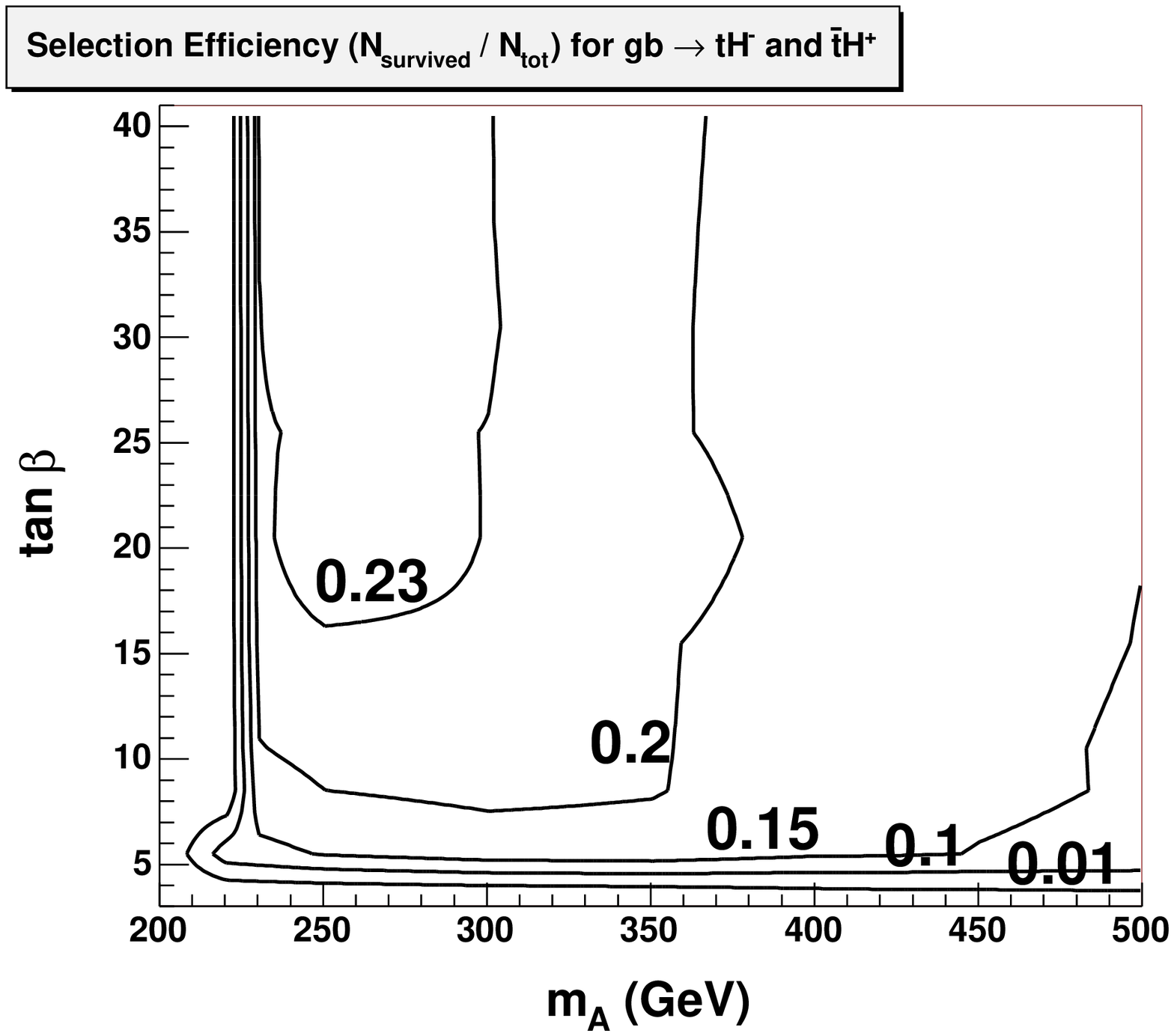}  &  \includegraphics[angle=0, scale= .30]{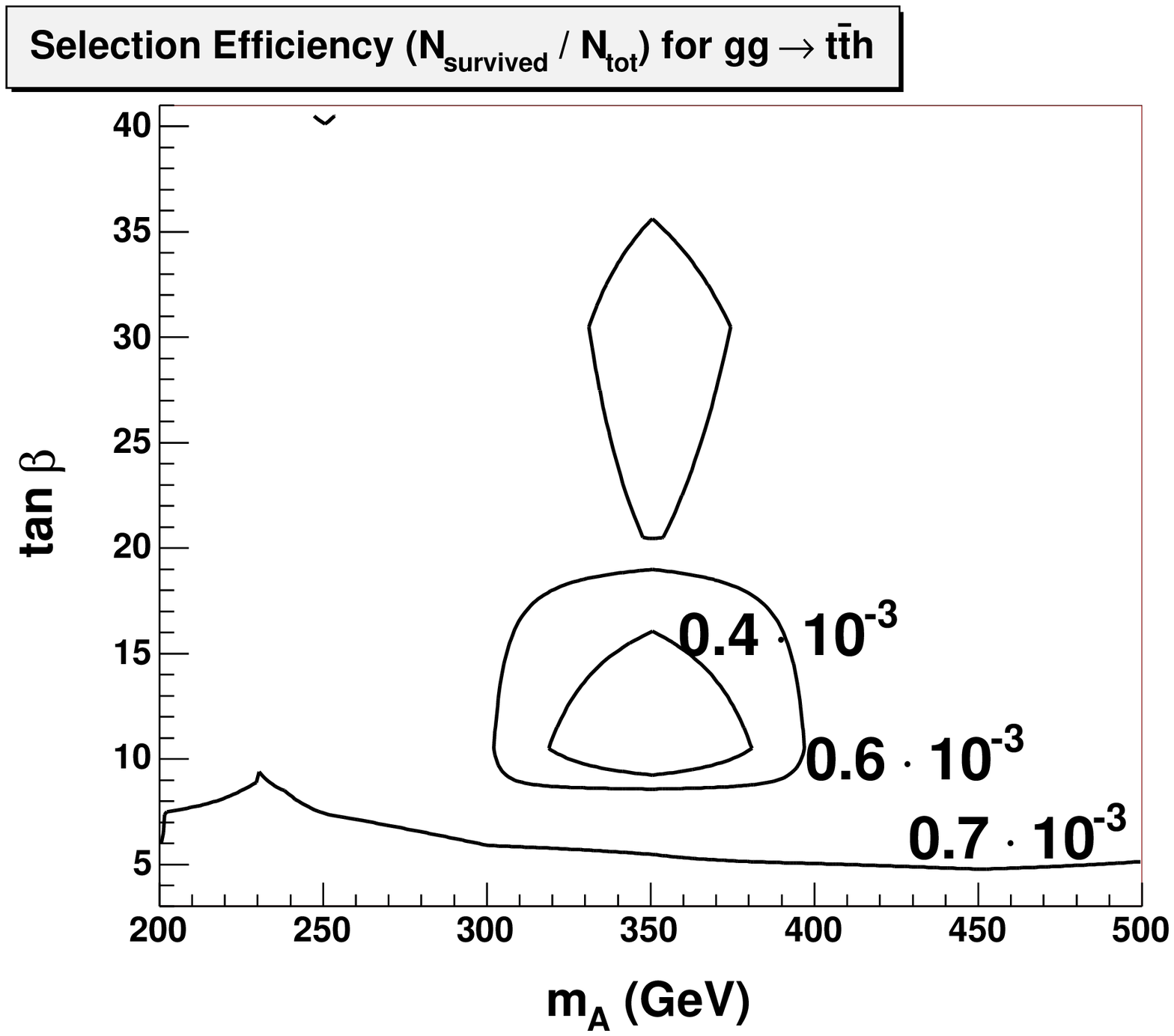} \\ [0.0cm]
\multicolumn{1}{c}{\mbox{\bf (c)}}                                &  \multicolumn{1}{c}{\mbox{\bf (d)}} \\ [-0.1cm]
\includegraphics[angle=0, scale= .30]{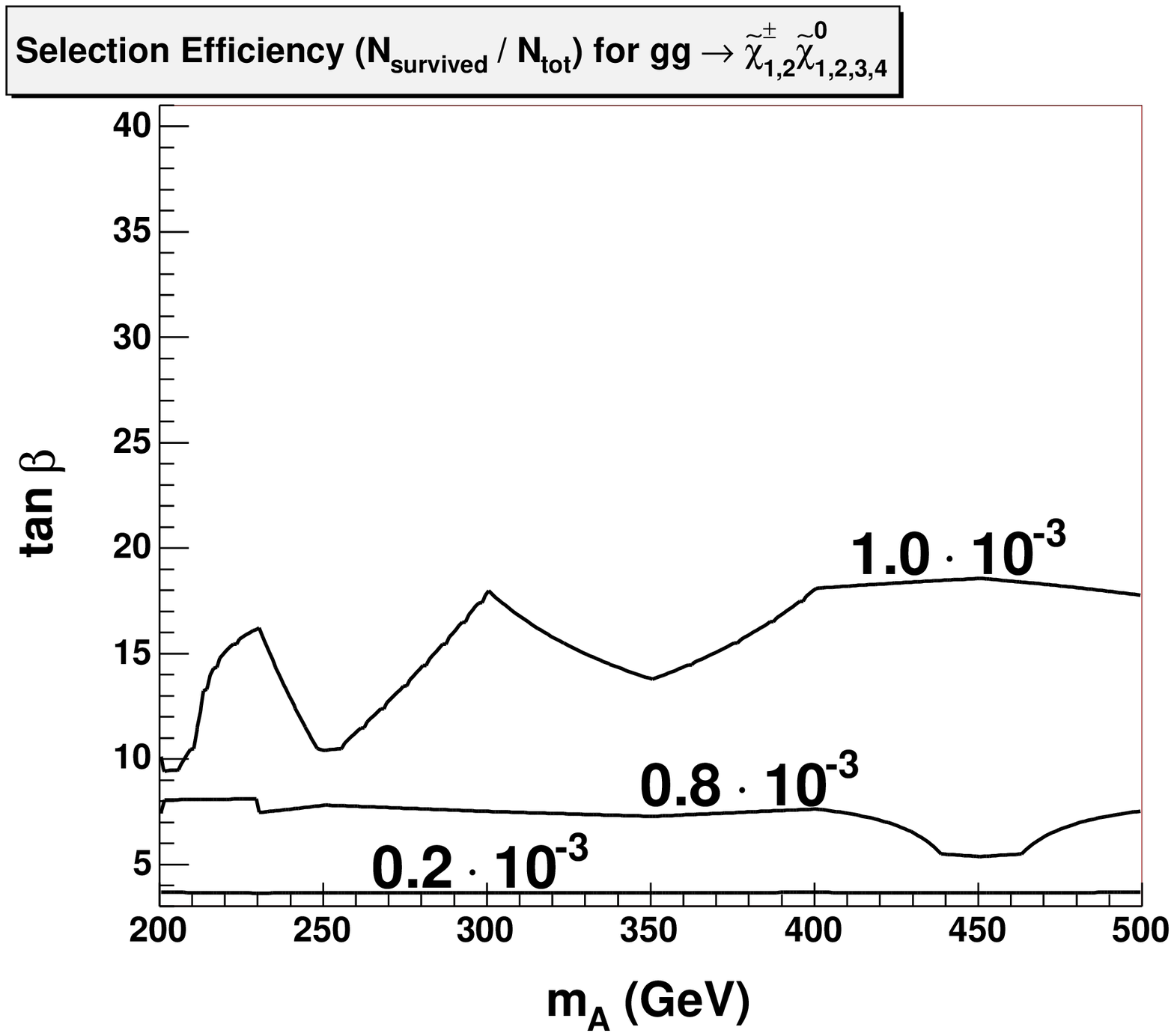}   &  \includegraphics[angle=0, scale= .30]{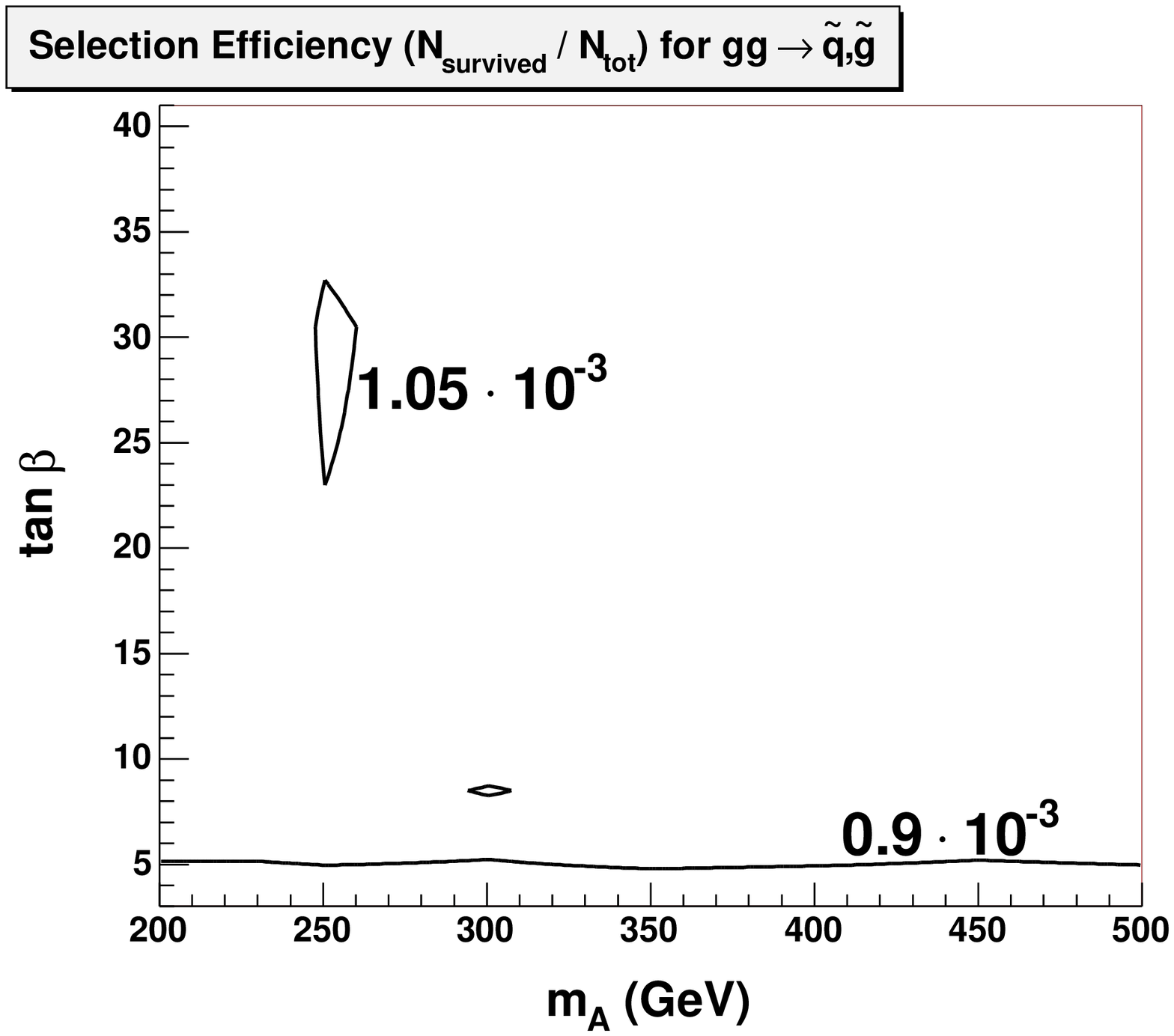} \\ [0.0cm]
\end{array}$
\end{center}
\caption{The contours show the selection efficiency for the signal ($gb \to tH^+, \bar{t}H^-$) and 
         for the SUSY background channel $gg \to t\bar{t}h$, $gg \to \tilde{\chi}\tilde{\chi}$
         and $gg \to \tilde{q}, \tilde{g}$
         in {\bf (a)}, {\bf (b)}, {\bf (c)} and  {\bf (d)} respectively.
         The efficiencies are calculated from simulated data taken from the ($\tan\beta, m_A$) values indicated in figure~\ref{fig:cs}a  
	 and in between the efficiency is calculated with linear interpolation.}
\label{fig:efficiencies}
\end{figure}
The selection efficiencies, i.e. the ratio of the number events selected by all cuts out of the total
number events, for the signal and the SUSY backgrounds are dependent on the ($\tan\beta$, $m_A$) value.
This is shown by the contours of constant selection efficiencies in figure~\ref{fig:efficiencies}.
The selection efficiencies for the SM background channels, not shown in figure~\ref{fig:efficiencies},
are, for all ($\tan\beta$, $m_A$) values, 5.4$\cdot$10$^{-5}$ and 3.6$\cdot$10$^{-4}$ for $gg \to t\bar{t}$ and $gg \to t\bar{t}Z$ respectively.

\begin{figure}[htp]   
\begin{center}
\includegraphics[angle=0, scale= .60]{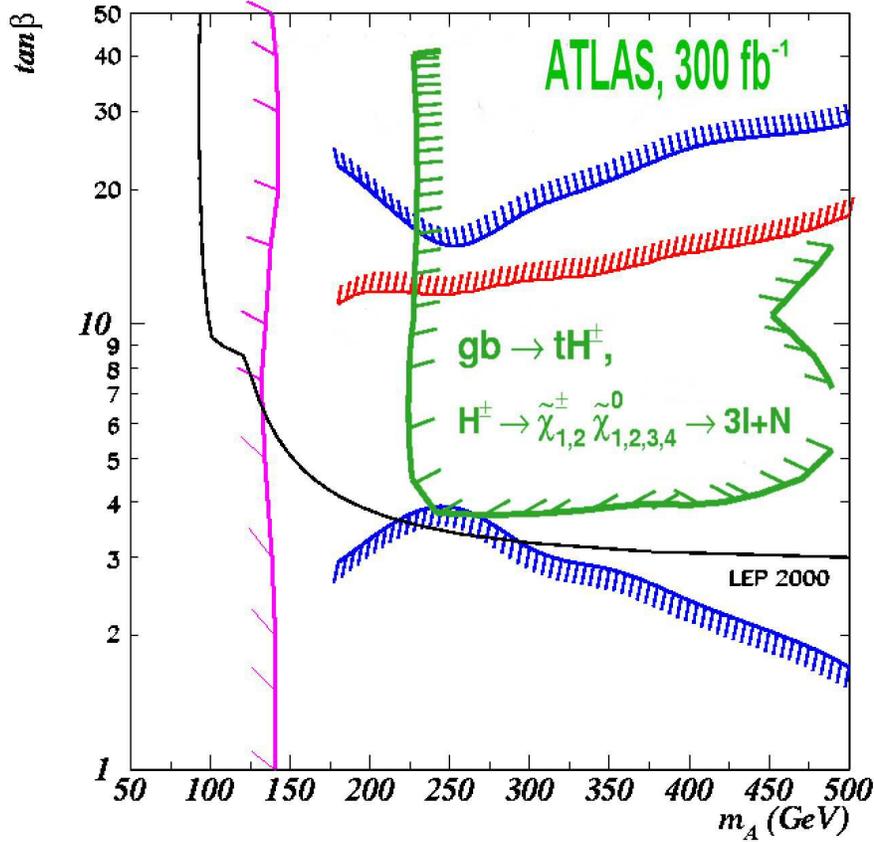}
\caption{The 5-$\sigma$ discovery contour for the 
         $H^{\pm}$~$\to$~$\tilde{\chi}^{\pm}_{1,2}\tilde{\chi}^0_{1,2,3,4}$~$\to$~$3\ell$~$+$~$N$ 
         channel for the parameter set defined in section~\ref{mssm} is shown in the $\tan\beta$ vs. $m_A$ plane. 
         The integrated luminosity is~300~fb$^{-1}$.}
\label{fig:scan}
\end{center}
\end{figure}
The selection efficiencies are used to also calculate the significance for different ($\tan\beta, m_A$) values.
This was done on a grid of points in the $\tan\beta$ vs. $m_A$ plane 
with $\tan\beta$ steps of 1 and $m_A$ steps of 25~GeV either directly from data at the data points 
(indicated by dots in figure~\ref{fig:cs}a) or by linear interpolation between the data points.
In figure~\ref{fig:scan} the contour where the significance reaches the value 5 is shown, 
superimposed on the previously introduced discovery contours shown in figure~\ref{fig:contour}.
The left edge of the potential discovery region 
more or less follows the $1\cdot 10^{-4}$ contour of 
branching ratio of $H^{\pm}$~$\to$~$\tilde{\chi}^{\pm}_{1,2}\tilde{\chi}^0_{1,2,3,4}$~$\to$~$3\ell+N$ 
shown in figure~\ref{fig:br}b. 
The right edge of the discovery region 
follow the pattern of the NLO cross section for the signal $gb \to tH^+, \bar{t}H^-$ shown in figure~\ref{fig:cs}a. 
\begin{figure}[tp]   
\begin{center}
\includegraphics[angle=0, scale= .60]{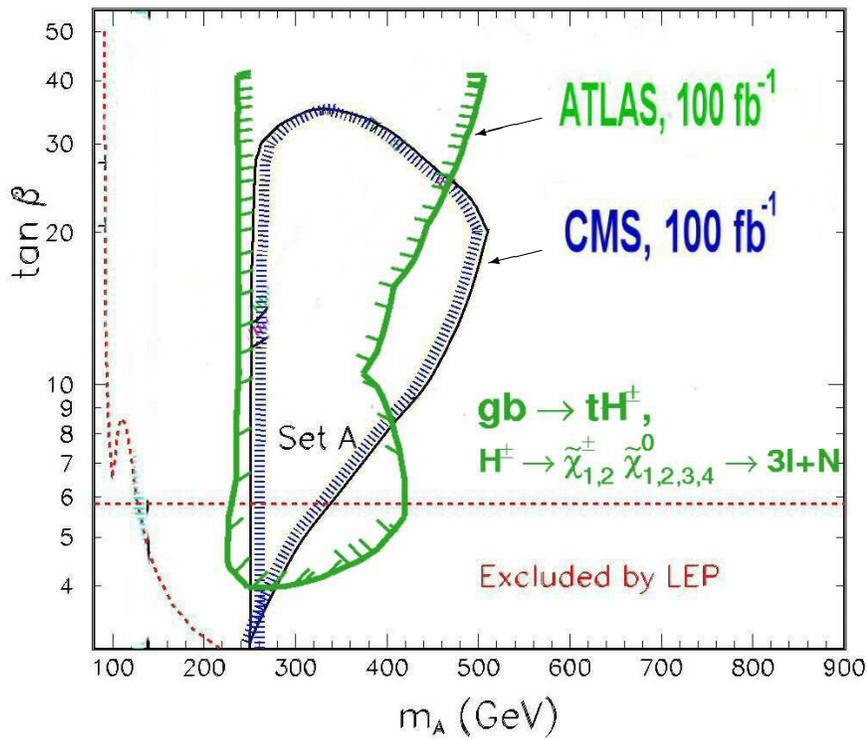}
\put(-125,235){\vector(-4,-1){25.}}
\put(-125,185){\vector(-4,-1){20.}}
\caption{The 5-$\sigma$ discovery contour for the 
         $H^{\pm}$~$\to$~$\tilde{\chi}^{\pm}_{1,2}\tilde{\chi}^0_{1,2,3,4}$~$\to$~$3\ell$~$+$~$N$ 
         channel for the parameter set defined in section~\ref{mssm} is shown in the $\tan\beta$ vs. $m_A$ plane, 
         superimposed over the corresponding result for CMS~\cite{CMS}. The integrated luminosity is~100~fb$^{-1}$.}
\label{fig:cmsandatlas}
\end{center}
\end{figure}

\section{Conclusions}
\label{conclusions}
With the specific SUSY parameter set chosen, the contour of 5-$\sigma$ significance for charged Higgs discovery through its decays to SUSY 
particles encloses the major part of the intermediate $\tan\beta$ region
not covered by the charged Higgs to SM particles decays as shown in figure~\ref{fig:contour}.
The result shows a discovery region about the same size as that obtained 
for CMS~\cite{CMS} using the same parameter set (the two regions are compared in 
figure~\ref{fig:cmsandatlas} where L~=~100~fb$^{-1}$).
However the regions for ATLAS and CMS differ in shape due to the difference in the  
selection criteria used, in the cross section assumptions (in the present analysis the NLO cross sections for 
$gb \to tH^+, \bar{t}H^-$ were used, whereas in~\cite{CMS} LO cross sections were used) and in the 
detectors used (described by ATLFAST for ATLAS and by CMSJET for CMS).

As can be seen in table~\ref{tab:res} there are about ten times more  $t\bar{t}$ and about 
three times more $\chi\tilde{\chi}$ events than the signal events present after the last cut.
Since no discriminating signature which helps to extract the signal from the remaining
background could be identified a counting experiment is performed.
However, in doing so, the assumption that the cross sections for the various background 
processes have been measured already is relied upon.

Many more MSSM parameter sets need to be analysed before more general conclusions can be made,
regarding the region in the parameter space where charged Higgs can be discovered.
However, as a preliminary result, this analysis shows that for a specific SUSY parameter set
SUSY decays of the charged Higgs bosons can be detected in ATLAS. 
This is especially interesting for the intermediate $\tan \beta$ region of $4$~$\lesssim$~$\tan\beta$~$\lesssim$~10, 
where SM decays of the charged Higgs cannot be detected.

\vspace{3 cm}

\section*{Acknowledgments}

This analysis has been performed within the framework of the ATLAS Collaboration.
We have made use of the physics analysis framework and tools which are the result of 
collaboration-wide efforts. 
The authors would like to acknowledge useful discussions with many colleagues in the Collaboration.
Further we would like to thank Stefano Moretti, Filip Moortgat and Mike Bisset 
for helpful advice and discussions. 
And we would also like to acknowledge Frank Paige for theoretical
discussions and for ISAJET input.
The MC samples were produced on the NorduGrid~\cite{GRID} and we thank the
NorduGrid team, in particular Mattias Ellert, for their support.

\newpage

\bibliographystyle{utphys}
\bibliography{hPlusToSusyPaper}

\providecommand{\href}[2]{#2}\begingroup\raggedright\begin{thebibliography}{10}

\bibitem{HIGGS}
{Higgs, Peter W.}, ``{Broken symmetries, massless particles and gauge
  fields},'' {\em {Phys. Lett.}} {\bf {12}} ({1964})
{132--133}.

\bibitem{SUSYSEARCH}
{Haber, Howard E. and Kane, Gordon L.}, ``{The Search for Supersymmetry:
  Probing Physics Beyond the Standard Model},'' {\em {Phys. Rept.}} {\bf {117}}
  ({1985})
{75}.

\bibitem{SUPER}
{Nilles, Hans Peter}, ``{Supersymmetry, Supergravity and Particle Physics},''
  {\em {Phys. Rept.}} {\bf {110}} ({1984})
{1}.

\bibitem{YANN}
{Assamagan, Ketevi Adikle and Coadou, Yann and Deandrea, Aldo}, ``{ATLAS
  discovery potential for a heavy charged {H}iggs boson},'' {\em {Eur. Phys. J.
  direct}} {\bf {C4}} ({2002}) {9},
\href{http://www.arXiv.org/abs/{hep-ph/0203121}}{{\tt {hep-ph/0203121}}}.

\bibitem{LEP}
{\bf {LEP Higgs Working Group for Higgs boson searches}} Collaboration, {},
  ``{Search for charged {H}iggs bosons: Preliminary combined results using
  {{LEP}} data collected at energies up to 209- {G}e{V}},''
\href{http://www.arXiv.org/abs/{hep-ex/0107031}}{{\tt {hep-ex/0107031}}}.

\bibitem{JOURNAL}
{\bf {Particle Data Group}} Collaboration, {Eidelman, S. and others}, ``{Review
  of particle physics},'' {\em {Phys. Lett.}} {\bf {B592}} ({2004})
{1}.

\bibitem{DENEGRI}
{Denegri, D. and others}, ``{Summary of the CMS discovery potential for the
  MSSM SUSY Higgses},''
\href{http://www.arXiv.org/abs/{hep-ph/0112045}}{{\tt {hep-ph/0112045}}}.

\bibitem{CAVALLI}
{D. Cavalli et al.}, ``{Search for $H^+ \to \tau \nu_{\tau}$ decays}.''
  {ATL-PHYS-94-053}, 1994.

\bibitem{TOPMASS}
{\bf {D0}} Collaboration, {Abazov, V. M. and others}, ``{A precision
  measurement of the mass of the top quark},'' {\em {Nature}} {\bf {429}}
  ({2004}) {638--642},
\href{http://www.arXiv.org/abs/{hep-ex/0406031}}{{\tt {hep-ex/0406031}}}.

\bibitem{KETEVI1}
{Ketevi Adikle Assamagan}, ``{The Charged Higgs in Hadronic Decays With the
  ATLAS Detector}.'' {ATL-PHYS-99-013}, 1999.

\bibitem{KETEVI2}
{Assamagan, K. A. and Coadou, Y.}, ``{The hadronic $\tau$ decay of a heavy
  $H^{\pm}$ in ATLAS},'' {\em {Acta Phys. Polon.}} {\bf {B33}} ({2002})
{707--720}.

\bibitem{HTBYUKAWA}
{Miller, D. J. and Moretti, S. and Roy, D. P. and Stirling, W. James},
  ``{Detecting heavy charged Higgs bosons at the LHC with four b quark tags},''
  {\em {Phys. Rev.}} {\bf {D61}} ({2000}) {055011},
\href{http://www.arXiv.org/abs/{hep-ph/9906230}}{{\tt {hep-ph/9906230}}}.

\bibitem{NILS}
{Assamagan, Ketevi A. and Gollub, Nils}, ``{The ATLAS discovery potential for a
  heavy charged Higgs boson in g g $\to$ t b H+- with H+- $\to$ t b},''
\href{http://www.arXiv.org/abs/{hep-ph/0406013}}{{\tt {hep-ph/0406013}}}.

\bibitem{CMS}
{Bisset, Mike and Moortgat, Filip and Moretti, Stefano}, ``{Trilepton + top
  signal from chargino neutralino decays of {{MSSM}} charged {H}iggs bosons at
  the {{LHC}}},'' {\em {Eur. Phys. J.}} {\bf {C30}} ({2003}) {419--434},
\href{http://www.arXiv.org/abs/{hep-ph/0303093}}{{\tt {hep-ph/0303093}}}.

\bibitem{CMS2}
{Bisset, Mike and Guchait, Monoranjan and Moretti, Stefano}, ``{Signatures of
  MSSM charged Higgs bosons via chargino neutralino decay channels at the
  LHC},'' {\em {Eur. Phys. J.}} {\bf {C19}} ({2001}) {143--154},
\href{http://www.arXiv.org/abs/{hep-ph/0010253}}{{\tt {hep-ph/0010253}}}.

\bibitem{HERWIG}
{Corcella, G. and others}, ``{{HERWIG 6}: {A}n event generator for hadron
  emission reactions with interfering gluons (including supersymmetric
  processes)},'' {\em {JHEP}} {\bf {01}} ({2001}) {010},
\href{http://www.arXiv.org/abs/{hep-ph/0011363}}{{\tt {hep-ph/0011363}}}.

\bibitem{HERWIGSUSY}
{Moretti, Stefano and Odagiri, Kosuke and Richardson, Peter and Seymour,
  Michael H. and Webber, Bryan R.}, ``{Implementation of supersymmetric
  processes in the {HERWIG} event generator},'' {\em {JHEP}} {\bf {04}}
  ({2002}) {028},
\href{http://www.arXiv.org/abs/{hep-ph/0204123}}{{\tt {hep-ph/0204123}}}.

\bibitem{ATLFAST}
{Richter--Was, Elzbieta and Froidevaux, Daniel and Poggioli, Luc}, ``{ATLFAST
  2.0, a fast simulation package for ATLAS}.'' {ATL-PHYS-98-131}, 1998.

\bibitem{TILMAN}
{Boos, Eduard and Plehn, Tilman}, ``{Higgs-boson production induced by bottom
  quarks},'' {\em {Phys. Rev.}} {\bf {D69}} ({2004}) {094005},
\href{http://www.arXiv.org/abs/{hep-ph/0304034}}{{\tt {hep-ph/0304034}}}.

\bibitem{ISAJET769}
{Baer, Howard and Paige, Frank E. and Protopopescu, Serban D. and Tata,
  Xerxes}, ``{ISAJET 7.69: A Monte Carlo event generator for $pp$, $\bar pp$,
  and $e^+e^-$ reactions},''
\href{http://www.arXiv.org/abs/{hep-ph/0312045}}{{\tt {hep-ph/0312045}}}.

\bibitem{TOP}
{Frixione, Stefano and Nason, Paolo and Webber, Bryan R.}, ``{Matching NLO QCD
  and parton showers in heavy flavour production},'' {\em {JHEP}} {\bf {08}}
  ({2003}) {007},
\href{http://www.arXiv.org/abs/{hep-ph/0305252}}{{\tt {hep-ph/0305252}}}.

\bibitem{ATLASTRIGGER}
{Hauser, R.}, ``{The ATLAS trigger system},'' {\em {Eur. Phys. J.}} {\bf {C34}}
  ({2004})
{s173--s183}.

\bibitem{SUSYMEAS}
{Poggioli, Luc; Polesello, G; Richter-Was, Elzbieta; S\"oderqvist, J},
  ``{Precision SUSY measurements with ATLAS for SUGRA point 5}.''
  {ATL-PHYS-97-111; ATL-GE-PN-111}, 1997.

\bibitem{GRID}
{Eerola, P. and others}, ``{The NorduGrid architecture and tools},'' {\em
  {ECONF}} {\bf {C0303241}} ({2003}) {MOAT003},
\href{http://www.arXiv.org/abs/{physics/0306002}}{{\tt {physics/0306002}}}.

\end{thebibliography}\endgroup

\end{document}